\documentclass[11pt,showpacs,preprint,preprintnumbers,nofootinbib,
superscriptaddress,amsmath,amssymb]{revtex4}

\usepackage{graphicx}
\usepackage{dcolumn}
\usepackage{bm}
\usepackage{amssymb}
\usepackage{amsmath}
\usepackage{epsfig}
\usepackage{color}
\usepackage{slashed}
\usepackage[hypertex]{hyperref}
\newcolumntype{I}{!{\vrule width 1.3pt}}
\begin{document} 

\title{
Bounds on the mass of doubly-charged Higgs bosons \\
in the same-sign diboson decay scenario
}

\preprint{UT-HET 091}

\author{Shinya Kanemura}
\email{kanemu@sci.u-toyama.ac.jp}
\affiliation{Department of Physics, University of Toyama, \\3190 Gofuku,
Toyama 930-8555, Japan}
\author{Mariko Kikuchi}
\email{kikuchi@jodo.sci.u-toyama.ac.jp}
\affiliation{Department of Physics, University of Toyama, \\3190 Gofuku,
Toyama 930-8555, Japan}
\author{Kei Yagyu}
\email{keiyagyu@ncu.edu.tw}
\affiliation{Department of Physics, National Central University,
\\Chungli 32001, Taiwan}
\author{Hiroshi Yokoya}
\email{hyokoya@sci.u-toyama.ac.jp}
\affiliation{Department of Physics, University of Toyama, \\3190 Gofuku,
Toyama 930-8555, Japan}

\begin{abstract}
A direct search for doubly-charged Higgs bosons $H^{\pm\pm}$
 is one of the most important probe in the Higgs Triplet Model, which is
 motivated by generation mechanisms of tiny neutrino masses.
There are two major decay modes of $H^{\pm\pm}$; i.e., the same-sign
 dilepton decay $H^{\pm\pm}\to\ell^\pm\ell^\pm$ and the same-sign
 diboson decay $H^{\pm\pm}\to W^{\pm(*)}W^{\pm(*)}$. 
For the case where the former decay mode is dominant, the lower limit on
 the mass of $H^{\pm\pm}$ has been set at about 400~GeV by ATLAS and CMS
 Collaborations.
On the other hand, for the case where the latter decay mode is dominant, 
 no dedicated search has been performed in the past.
By taking into account characteristic signals of $H^{\pm\pm}$ in the
 diboson decay scenario at LEP and the LHC experiments, we find that the
 lower mass bound of 60-68~GeV can be obtained using
 the same-sign dilepton search performed by ATLAS Collaboration with
 4.7~fb$^{-1}$ data at the collision energy of 7~TeV. 
We also show that the limit can be extended up to about
 85-90~GeV, assuming the integrated luminosity of 20~fb$^{-1}$ and 7~TeV
 for the collision energy.
We give detailed explanations on the decay properties of $H^{\pm\pm}$
 for relatively small mass cases and also on production cross sections of
 $H^{\pm\pm}$ at the next-to-leading order of QCD at the LHC.
\end{abstract}

\pacs{12.60.Fr, 14.80.Fd}

\maketitle
\newpage

\section{Introduction}

In 2012, a Higgs boson was discovered at the CERN Large Hadron Collider
(LHC)~\cite{LHC_Higgs,LHC_Higgs2}.
Its observed properties are consistent with the prediction in the
standard model (SM) within the current experimental
uncertainties~\cite{Higgs_property}.
In addition, so far, no report has been delivered to us for discovery of
the other new particles.
Therefore, it has been found that the SM is a good description for
particle physics at the scale of hundred GeV, not only in the gauge
interactions but also in the sector of electroweak symmetry breaking.

Although the Higgs boson has been discovered and its property has turned
out to be SM-like, we know nothing about the structure of the Higgs
sector. 
In fact, the minimal Higgs sector with one isospin doublet scalar field
is just an assumption without any theoretical principle.
Thus, it is natural to consider a possibility that the Higgs sector
takes a non-minimal form with additional isospin multiplet scalar
fields, such as extra singlet, doublet, triplet and so on.
Most of these non-minimal Higgs sectors can explain current experimental
data as well.
Furthermore, these extended Higgs sectors are often introduced in the
context of new physics models which try to explain the phenomena beyond
the SM; i.e., neutrino masses, dark matter, and baryogenesis.
Therefore, it is very important to experimentally explore the
possibility of extended Higgs sectors. 
We then may be able to discriminate new physics models from the property
of the Higgs sector.

For example, extended Higgs sectors with multi-doublet scalar fields
are introduced in supersymmetric extensions of the SM.
They are also motivated to introduce an additional source of CP
violation~\cite{cp}, and to realize the strong 1st-order phase
transition~\cite{1opt}, both of which 
are required to have successful electroweak baryogenesis~\cite{ewbg}. 
Singlet scalar fields are often introduced in models with
the spontaneously broken $B-L$ gauge symmetry~\cite{b-l}.
The Higgs sector with a complex triplet scalar field appears in models
that can explain neutrino masses via the seesaw mechanism~\cite{typeII}.
Tiny neutrino masses can also be explained via the loop-induced 
effects of extended scalar sectors~\cite{Zee,KNT,Ma,AKS}.
Extended scalar sectors with a discrete symmetry such as $Z_2$ can
provide a candidate of dark matter~\cite{z2_d,z2_s}.

We here focus on the Higgs Triplet Model (HTM)~\cite{typeII}.
Its Higgs sector is composed of an isospin doublet Higgs field with a
hypercharge\footnote{%
We adopt the notation of $Y$ as $Q=T^3+Y$, where $Q$ is the electric
charge and $T^3$ is the third component of the isospin.}
$Y=1/2$ and an isospin triplet Higgs field with $Y=1$.
In this model, Majorana masses of neutrinos are generated via new
Yukawa interactions among the left-handed lepton doublets and the Higgs
triplet field; $(m_\nu)_{ij} \propto h_{ij}\,v_\Delta$, where $v_\Delta$
is the vacuum expectation value (VEV) of the triplet field, and $h_{ij}$
is a matrix in the Lagrangian for the Yukawa interactions. 

One of the most characteristic features of the HTM is the existence of
doubly-charged Higgs bosons $H^{\pm\pm}$, 
in addition to the other additional Higgs bosons; i.e.,
singly-charged $H^{\pm}$, CP-even $H$ and CP-odd $A$ Higgs bosons.
The discovery of $H^{\pm\pm}$ at collider experiments is the direct
evidence of the HTM. 
Production of these bosons at collider experiments has been studied in
Refs.~\cite{AKY,HTM_LHC,neutrino1,neutrino2,AA,AC,Han,Perez,trilepton,%
Azuelos:2004dm,ChunSharma,STY,Nomura,KYY,delAguila:2013mia,Kang:2014jia,%
Godfrey:2010qb,Cascade,Dutta:2014dba,HTM_mass_diff,Sugiyama,Sugiyama2}.
For the decay of $H^{\pm\pm}$, there are three sources; i.e., the Yukawa
interactions with left-handed lepton doublets, electroweak gauge
interactions in the gauge-gauge-scalar type, and those in the
gauge-scalar-scalar type. 
They cause the same-sign dilepton decay $H^{\pm\pm}\to \ell^\pm
\ell^\pm$, the same-sign diboson decay $H^{\pm\pm}\to W^\pm W^\pm$ and
the cascade decay $H^{\pm\pm}\to H^{\pm}W^{\pm}$, respectively\footnote{%
In principle, $H^{\pm\pm}\to H^\pm H^\pm$ decay occurs
via the scalar triple couplings, if there is a large mass difference
between $H^{\pm\pm}$ and $H^\pm$. 
However, such a situation is severely constrained by electroweak
precision measurements~\cite{AKKY_full}.}.

Although the dominant decay mode of $H^{\pm\pm}$ is determined by
parameters in the model, the dilepton decay scenario has been considered
as the most promising one for
discovery~\cite{HTM_LHC,trilepton,neutrino1,neutrino2,AC,Perez,STY}, 
because of its cleanness for the detection at colliders.
It is also quite appealing that the structure of neutrino mass matrix
can be directly tested by measuring the dileptonic branching ratios of
$H^{\pm\pm}$~\cite{neutrino1,neutrino2,Perez} and
$H^\pm$~\cite{Perez,Akeroyd:2013kga}, because the branching ratios are
predominantly determined by the neutrino Yukawa couplings.
In this scenario, a sharp peak in the invariant mass distribution of 
the same-sign dilepton is the characteristic signal of $H^{\pm\pm}$.
The experimental searches for $H^{\pm\pm}$ in the same-sign dilepton
events have been performed at LEP~\cite{LEP}, HERA~\cite{HERA},
Tevatron~\cite{CDF,D0} and the LHC~\cite{400GeV_ATLAS,400GeV_CMS}.
Assuming that the branching ratio of $H^{\pm\pm}$ decay into
$\mu^{\pm}\mu^{\pm}$ is 100\%, the strongest lower bound on the mass of
doubly-charged Higgs bosons has been obtained as 459~GeV at the
LHC~\cite{400GeV_CMS}. 
Current bounds have also been set at around 400~GeV in 
several benchmark points for the structure of the neutrino mass
matrix~\cite{400GeV_CMS}.

In this paper, we discuss the direct searches for $H^{\pm\pm}$ in the
diboson decay scenario, where $H^{\pm\pm}$ predominantly
decay into same-sign W bosons, at the past, current and future
collider experiments, such as LEP, the LHC with 7-8~TeV run and
13-14~TeV run.
The same-sign diboson decay scenario is equally important to the
same-sign dilepton decay scenario in the HTM.
Collider phenomenology for this decay mode has been studied in
Refs.~\cite{Han,Nomura,KYY,Kang:2014jia}.
In Ref.~\cite{KYY}, the lower limit on the mass of $H^{\pm\pm}$ has been
derived by using the same-sign dilepton events collected by ATLAS
Collaboration at the LHC with 7~TeV and 4.7~fb$^{-1}$
data~\cite{ATLAS}.
Up to our knowledge, this is the first analysis for the constraints on
$H^{\pm\pm}$ in the diboson scenario.
The aim of the present paper is to explain details of the analysis done in
Ref.~\cite{KYY}, and to make update on the results by including the QCD
correction to the production cross sections.

This paper is organized as follows. 
In Section II, we briefly review the HTM.
After we define the mass eigenstates for the Higgs bosons, we derive the
Yukawa interaction and the gauge interaction for the triplet-like Higgs
bosons at the tree level. 
In Section III, we give expressions for the decay rates of $H^{\pm\pm}$
in the all three decay modes.
Partial decay widths of $H^{\pm\pm}$ are evaluated with
particular attention to the case of relatively small masses where one or
both of the W bosons are forced off-shell.
We then show the phase diagram indicating the main decay mode of
$H^{\pm\pm}$.
Next, we evaluate the cross section of $H^{\pm\pm}$ productions at the
LHC in the leading order (LO) and the next-to-leading order (NLO) of
QCD. 
In Section IV, we exhibit constraints on the mass of $H^{\pm\pm}$
in the diboson decay scenario at the LEP experiments and also at the
LHC.
At the LEP~I experiment, the lower limit on the mass of $H^{\pm\pm}$ can
be obtained from the total width of the Z boson.
We also evaluate the expected number of events for the various final
states in the process of $e^+e^-\to H^{++}H^{--}$ at the LEP~II
experiment. 
We then discuss the mass bound on $H^{\pm\pm}$ by using the current LHC
limit on the cross section for anomalous production of same-sign
dileptons. 
Section V is devoted to our conclusion.
In Appendix, the cross sections for $H^{\pm\pm}$ production at the LHC
with various collision energies are collected for reader's convenience.

\section{The Higgs Triplet Model}

The scalar sector of the HTM is composed of the isospin doublet field
$\Phi$ with hypercharge $Y=1/2$ and the triplet field $\Delta$ with
$Y=1$. 
The most general form of the Higgs potential under the gauge symmetry is
written as 
\begin{align}
V(\Phi,\Delta)&=
m^2\Phi^\dagger\Phi+M^2\text{Tr}(\Delta^\dagger\Delta)
+\left[\mu\Phi^Ti\tau_2\Delta^\dagger \Phi+\text{h.c.}\right]\notag\\
&+\lambda_1(\Phi^\dagger\Phi)^2
+\lambda_2\left[\text{Tr}(\Delta^\dagger\Delta)\right]^2
+\lambda_3\text{Tr}\left[(\Delta^\dagger\Delta)^2\right]
+\lambda_4(\Phi^\dagger\Phi)\text{Tr}(\Delta^\dagger\Delta)
+\lambda_5\Phi^\dagger\Delta\Delta^\dagger\Phi,
 \label{pot_htm}
\end{align}
where all the parameters are taken to be real without loss of
generality~\cite{Dey}. 
The Higgs fields can be parameterized as
\begin{align}
\Phi=\left(
\begin{array}{c}
\phi^+\\
\phi^0
\end{array}\right),\quad \Delta =
\left(
\begin{array}{cc}
\frac{\Delta^+}{\sqrt{2}} & \Delta^{++}\\
\Delta^0 & -\frac{\Delta^+}{\sqrt{2}} 
\end{array}\right), 
\end{align}
where the neutral components are expressed as
\begin{align}
\phi^0=\frac{1}{\sqrt{2}}(\phi_R^0+v_\phi+i\phi_I^0),\quad 
\Delta^0=\frac{1}{\sqrt{2}}(\Delta_R^0+v_\Delta+i\Delta_I^0). 
\end{align}
The VEVs of the doublet and triplet Higgs fields are denoted by 
$v_\phi$ and $v_\Delta$, respectively.  
They are related to the Fermi constant $G_F$ by $v^2 \equiv
v_\phi^2+2v_\Delta^2=(\sqrt{2}G_F)^{-1}$. 
The non-zero $v_\Delta$ deviates the electroweak rho parameter from
unity at the tree level; 
\begin{align}
 \rho \equiv
 \frac{m_W^2}{m_Z^2\cos^2\theta_W}=
 \frac{1+\frac{2v_\Delta^2}{v_\phi^2}}{1+\frac{4v_\Delta^2}{v_\phi^2}},
 \label{rho_triplet}
\end{align}
where $m_W$, $m_Z$ and $\theta_W$ are the W boson mass, the Z boson mass
and the weak mixing angle, respectively. 
Since the experimental value of the rho parameter is close to unity; i.e.,
$\rho_{\text{exp}}=1.0004^{+0.0003}_{-0.0004}$~\cite{PDG}, 
$v_\Delta$ has to be less than about 3.5~GeV at the 95\% confidence
level~(CL). 

Mass eigenstates in the doubly-charged states ($H^{\pm\pm}$) purely come
from $\Delta$; i.e., $H^{\pm\pm}=\Delta^{\pm\pm}$.
For the other scalar bosons, mass eigenstates are defined by
introducing the following orthogonal transformations;
\begin{align}
\left(
\begin{array}{c}
\phi_R^0\\
\Delta_R^0
\end{array}\right)&=
\left(
\begin{array}{cc}
\cos \alpha & -\sin\alpha \\
\sin\alpha   & \cos\alpha
\end{array}
\right)
\left(
\begin{array}{c}
h\\
H
\end{array}\right), \quad
\left(
\begin{array}{c}
\phi^\pm\\
\Delta^\pm
\end{array}\right)=
\left(
\begin{array}{cc}
\cos \beta & -\sin\beta \\
\sin\beta   & \cos\beta
\end{array}
\right)
\left(
\begin{array}{c}
G^\pm\\
H^\pm
\end{array}\right),\notag\\ 
\left(
\begin{array}{c}
\phi_I^0\\
\Delta_I^0
\end{array}\right) &=
\left(
\begin{array}{cc}
\cos \beta' & -\sin\beta' \\
\sin\beta'   & \cos\beta'
\end{array}
\right)
\left(
\begin{array}{c}
G^0\\
A
\end{array}\right),
\label{mixing1}
\end{align}
where mixing angles, $\alpha$, $\beta$ and $\beta'$ are given by 
\begin{align}
\tan2\alpha
 &=\frac{v_\Delta}{v_\phi}
 \frac{2v_\phi^2(\lambda_4+\lambda_5)-4M_\Delta^2}
 {2v_\phi^2\lambda_1-M_\Delta^2-2v_\Delta^2(\lambda_2+\lambda_3)}, \quad
\tan\beta=\frac{\sqrt{2}v_\Delta}{v_\phi},\quad \tan\beta' =
 \frac{2v_\Delta}{v_\phi},
 \label{Eq:mixing_angles} 
\end{align}
with 
\begin{align}
M_\Delta^2\equiv \frac{v_\phi^2\mu}{\sqrt{2}v_\Delta}. 
\end{align}
In Eq.~(\ref{mixing1}), $G^\pm$ and $G^0$ are the Nambu-Goldstone bosons
which are absorbed into the longitudinal component of W and Z bosons,
respectively. 
Because all the mixing angles given in Eq.~(\ref{Eq:mixing_angles}) are
quite small due to $v_\Delta/v_\phi\ll 1$, 
$H^\pm$, $A$ and $H$ are mostly composed of the triplet Higgs field. 
We thus call these scalars ($H^\pm$, $A$, $H$ and $H^{\pm\pm}$) as the
triplet-like Higgs bosons. 
On the other hand, by the same reason, $h$ can be regarded as the
SM-like Higgs boson, because it mainly comes from the doublet Higgs
field. 
By neglecting $\mathcal{O}(v_\Delta^2)$ terms, 
the masses of these physical Higgs bosons are given in a good
approximation by 
\begin{align}
&m_{H^{\pm\pm}}^2 \simeq M_\Delta^2-\frac{\lambda_5}{2}v^2,\quad 
m_{H^\pm}^2 \simeq M_\Delta^2-\frac{\lambda_5}{4}v^2,\quad 
m_A^2 \simeq m_H^2 \simeq M_\Delta^2,\\
&m_h^2 \simeq 2\lambda_1v^2. 
\end{align}
Thus, it can be observed that there are relationships among the masses
of triplet-like Higgs bosons~\cite{AKY,HTM_mass_diff}; i.e.,
$m_{H^{\pm\pm}}^2-m_{H^{\pm}}^2\simeq m_{H^{\pm}}^2-m_A^2$ and
$m_A^2\simeq m_H^2$. 
From these relations, three patterns of the mass spectrum arise.
First two patterns are $m_{A}>m_{H^\pm}>m_{H^{\pm\pm}}$ in the case with
$\lambda_5>0$ and $m_{H^{\pm\pm}}>m_{H^\pm}>m_A$ in the case with
$\lambda_5<0$.
In the special case with $\lambda_5=0$, all the triplet-like Higgs
bosons degenerate in mass. 

The kinetic term of the Lagrangian for the Higgs fields is given by 
\begin{align}
\mathcal{L}_{\text{kin}}=|D_\mu\Phi|^2+\text{Tr}
 \left[(D_\mu\Delta)^\dagger(D^\mu\Delta)\right], 
\end{align}
where the covariant derivatives are defined as
\begin{align}
D_\mu\Phi=\left(\partial_\mu -i\frac{g}{2}\tau^aW^a_\mu
 -i\frac{g'}{2}B_\mu\right)\Phi,\quad D_\mu\Delta=\partial_\mu
 \Delta-i\frac{g}{2}\left[\tau^aW^a_\mu,\Delta\right] -ig'B_\mu\Delta. 
\end{align}
From the above Lagrangian, Higgs-gauge-gauge type vertices are derived.
Coefficients of the vertices for the triplet-like Higgs bosons are given
as follows:
\begin{align}
&(H^{\pm\pm}W^\mp_\mu W^\mp_\nu):
-gm_W\sin\beta g_{\mu\nu},\quad 
(H^{\pm}W^\mp_\mu Z_\nu):
-\frac{g}{\cos\theta_W}m_W\sin\beta\cos\beta
 g_{\mu\nu} \notag \\
&(HW^\pm_\mu W^\mp_\nu):
-gm_W(\cos\beta\sin\alpha-\sqrt{2}\sin\beta\cos\alpha)g_{\mu\nu},\notag\\ 
&(HZ_\mu Z_\nu): -\frac{g}{\cos\theta_W}m_Z
(\cos\beta'\sin\alpha-2\sin\beta'\cos\alpha)g_{\mu\nu}.
\label{Eq:gauge_coupling}
\end{align} 
We note that according to Eq.~(\ref{Eq:mixing_angles}), all the
couplings are proportional to $v_\Delta/v$.

Next, we introduce the Yukawa interaction terms with the triplet field. 
Left-handed lepton doublet fields $L_L$ can couple to the triplet Higgs
field by, 
\begin{align}
\mathcal{L}_Y&=h_{ij}\overline{L_L^{ic}}i\tau_2\Delta L_L^j+\text{h.c.} 
\end{align}
If we extract the VEV in the neutral component of the triplet field, we
find a Majorana mass term for neutrinos~\cite{typeII},
\begin{align}
(m_\nu)_{ij}=\sqrt{2}h_{ij}v_\Delta.  \label{Eq:mn}
\end{align}
Couplings of the Yukawa interactions among the triplet-like Higgs bosons
and leptons are expressed in terms of $v_\Delta$ and the neutrino mass
matrix, $(m_{\nu})_{ij}$ with the use of Eq.~(\ref{Eq:mn}) as follows: 
\begin{align}
&(H^{++}\ell_i^- \ell_j^-): -\frac{(m_\nu)_{ij}}{\sqrt{2}v_\Delta}P_L
,\quad
(H^+\ell_i^- \nu_j):-\frac{(m_\nu)_{ij}}{v_\Delta}\cos\beta P_L   \notag\\
&(H\nu_i\nu_j):\frac{(m_\nu)_{ij}}{2v_\Delta}\cos\alpha P_L,\quad 
(A\nu_i\nu_j):i\frac{(m_\nu)_{ij}}{2v_\Delta}\cos\beta' P_L, 
\label{Eq:lepton_coupling}
\end{align}
where $P_L$ is the left-handed projection operator $(1-\gamma_5)/2$. 
From Eqs.~(\ref{Eq:gauge_coupling}) and (\ref{Eq:lepton_coupling}), we
see that the gauge (Yukawa) coupling constants are enhanced (suppressed)
as $v_\Delta$ gets increased. 
These features are important to understand the decay property of the
triplet-like Higgs bosons which is discussed in the next section. 

We note that the interaction terms between quarks and triplet-like Higgs
bosons except $H^{\pm\pm}$ are induced from the Yukawa interaction for
the doublet Higgs field $\Phi$ via the small mixing denoted by $\alpha$,
$\beta$ and $\beta'$~\cite{Perez,AKY}.

\section{Decay and Production of $H^{\pm\pm}$}

In this section, we discuss the decay and production of $H^{\pm\pm}$.
For the decay of $H^{\pm\pm}$, we present the decay rates for all the
three decay modes.
Especially, we discuss the diboson decay mode in detail, focusing on the
cases where one or both of the W bosons are forced off-shell.
For the production of $H^{\pm\pm}$ at the LHC, we evaluate the
cross sections in the LO and the NLO in QCD.
We estimate the uncertainties of theoretical calculations by taking into
account the scale ambiguity and the uncertainty from parton distribution
functions (PDFs).

\subsection{Decay branching ratio of $H^{\pm\pm}$} 

The decay properties of $H^{\pm\pm}$ strongly depend on $v_\Delta$ and
the mass spectrum of the triplet-like Higgs bosons. 
For the case where $H^{\pm\pm}$ are the lightest among all the
triplet-like Higgs bosons; i.e., $m_{A/H} \geq m_{H^\pm}\geq
m_{H^{\pm\pm}}$, the same sign dilepton decay $H^{\pm\pm}\to
\ell^\pm\ell^\pm$ and the same sign diboson decay $H^{\pm\pm}\to W^{\pm
(*)} W^{\pm (*)}$ are possible. 
On the other hand, for the case where $H^{\pm\pm}$ are the heaviest,
$m_{H^{\pm\pm}}> m_{H^\pm} > m_{A/H} $, another cascade-type decay
$H^{\pm\pm}\to W^{\pm(*)}H^\pm$ is also possible.

For the diboson decay, in the case with $m_{H^{\pm\pm}}\geq 2m_W$, 
the tree level decay rate is given by
\begin{align}
 \Gamma(H^{\pm\pm}\to W^\pm W^\pm) &=
 \frac{\sqrt{2}G_F\sin^2\beta}{8\pi}
 m_{H^{\pm\pm}}^3
 \left(1-4\frac{m_W^2}{m_{H^{\pm\pm}}^2}
 +12\frac{m_W^4}{m_{H^{\pm\pm}}^4}\right)
 \sqrt{1-\frac{4m_W^2}{m_{H^{\pm\pm}}^2}}.
 \label{Eq:diboson}
\end{align}
Furthermore, the branching ratio for four-fermion final-states
is simply given by multiplying the decay branching ratio of the W
bosons; i.e., 
\begin{align}
\mathcal{B}(H^{\pm\pm}\to 4f)= \mathcal{B}(H^{\pm\pm}\to W^\pm
 W^\pm)\times \mathcal{B}(W\to f\bar{f}')\times \mathcal{B}(W\to
 f''\bar{f}''').
\label{Eq:on-shell}
\end{align} 
On the other hand, in the case with $m_{H^{\pm\pm}}<2m_W$, at least
one of the W bosons is forced off-shell, and the decay rate
given in Eq.~(\ref{Eq:diboson}) is no longer valid. 
Thus, the branching ratio of $H^{\pm\pm}$ into four-fermion
final-states is not simply described by Eq.~(\ref{Eq:on-shell}).
In order to clarify how the difference in the decay rate of $H^{\pm\pm}$
appears in the case with the off-shell W boson(s), we first consider the
decay process of $H^{\pm\pm}$ into the four-lepton final-states; 
\begin{align}
 H^{\pm\pm}\to W^{\pm(*)}W^{\pm(*)}\to\ell^\pm\ell^{\pm}\nu\nu.
\label{Eq:llvv}
\end{align}
We can divide the decay modes into two cases; one is the same-flavour
(s.f.) dilepton mode such as $e^\pm e^\pm$, $\mu^\pm\mu^\pm$, and the
other is the different-flavour (d.f.) dilepton mode such as
$e^\pm\mu^\pm$.
\begin{figure}[t]
 \begin{center}
  \includegraphics[width=0.5\textwidth]{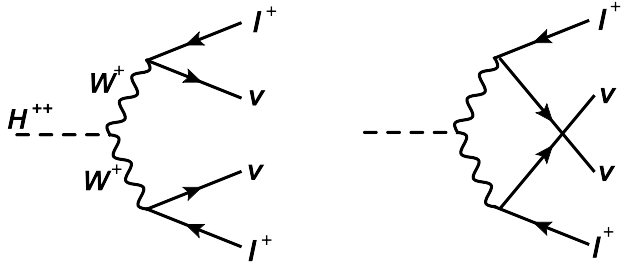}
  \caption{Feynman diagrams of the $H^{++}\to
  \ell^+\ell^+\nu_\ell\nu_\ell$ decay.
  Both the two diagrams contribute for the same-flavour dilepton cases,
  while only one diagram contributes for the different-flavour dilepton
  cases. 
}\label{fig:llvv} 
 \end{center}
\end{figure}
In the s.f.\ dilepton decay, two Feynman diagrams drawn in
Fig.~\ref{fig:llvv} contribute, while only one Feynman diagram 
contributes in the d.f.\ dilepton decay. 
For each case, the partial decay width is calculated as 
\begin{align}
 & \Gamma(H^{\pm\pm}\to \ell^\pm\ell^\pm\nu\nu) = 
 \Gamma_{\rm s.f.} \equiv \frac{g^8v_\Delta^2}{m_{H^{\pm\pm}}}
 \frac{1}{4}\int d\Phi_4
 \left|\Delta_{13}\Delta_{24}+\Delta_{14}\Delta_{23}\right|^2 
 (2p_1\cdot p_2)(2p_3\cdot p_4), \label{sfdecay}  \\
 & \Gamma(H^{\pm\pm}\to \ell^\pm\ell'^\pm\nu\nu') = 
 \Gamma_{\rm d.f.} \equiv \frac{g^8v_\Delta^2}{m_{H^{\pm\pm}}}
 \int d\Phi_4
 \left|\Delta_{13}\Delta_{24}\right|^2 (2p_1\cdot p_2)(2p_3\cdot p_4),
\end{align}
where $p^\mu_{i}$ with $i=1,...,4$ are the four momenta of the
final-state leptons in the order of the last term in
Eq.~(\ref{Eq:llvv}),
$\Delta_{ij}=[(p_i+p_j)^2-m_W^2+im_W\Gamma_W]^{-1}$, and $\int d\Phi_4$
denotes full phase-space integration over the four-body final-state. 
We neglect the mass of leptons. 
The difference between the two widths exists only in the interference
term in Eq.~(\ref{sfdecay}).
Similarly, the partial decay width for the $\ell\nu jj$ channel is given
by $\Gamma(H^{\pm\pm}\to \ell^\pm\nu jj) = 6\Gamma_{\rm d.f.}$ for each
lepton flavour, where a jet $j$ includes $u,~d,~c$ and $s$ quarks and
those anti-particles. 
In addition, the partial decay width for the $jjjj$ channel is given by
$\Gamma(H^{\pm\pm}\to jjjj) = 6\Gamma_{\rm s.f.} + 15\Gamma_{\rm d.f.}$. 
In total, the sum of the decay width through $H^{\pm\pm}\to
W^{\pm(*)}W^{\pm(*)}$ is given by $\Gamma_{H^{\pm\pm}}=9\Gamma_{\rm
s.f.}+36\Gamma_{\rm d.f.}$.
We note that in the case with $m_{H^{\pm\pm}}>2m_W$ where both the W
bosons can be on-shell, $\Gamma_{\rm s.f.}=\Gamma_{\rm d.f.}/2$ is a good
approximation by neglecting the interference term, and the branching
ratios reduce to the product of branching ratios of the W bosons as in
Eq.~(\ref{Eq:on-shell}). 

\begin{figure}[t]
 \begin{center}
  \includegraphics[width=0.5\textwidth]{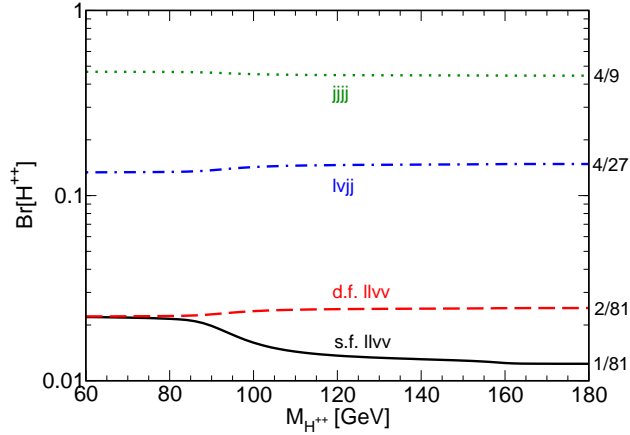}
  \caption{Branching ratios of $H^{\pm\pm}$ into $jjjj$, $\ell^\pm\nu jj$,
  same-flavour and different-flavour $\ell^\pm\ell^\pm\nu\nu$ modes as a
  function of $m_{H^{\pm\pm}}$.
  In this plot, only the $H^{\pm\pm}\to W^{\pm(*)}W^{\pm(*)}$ mode is
  taken into account. 
  }\label{fig:Br4l}
 \end{center}
\end{figure}

In Fig.~\ref{fig:Br4l}, we plot the branching ratios for the decay of
$H^{\pm\pm}$ into various four-fermion final states, such as $jjjj$,
$\ell^\pm\nu jj$ and $\ell^\pm\ell^\pm\nu\nu$ with s.f.\ or d.f.\
leptons, as a function of $m_{H^{\pm\pm}}$. 
Notice that we neglect the dilepton decay and cascade decay
channels here.
It is found that the branching ratio of the s.f.\
$\ell^\pm\ell^\pm\nu\nu$ decay mode is enhanced by 80\% for
$m_{H^{\pm\pm}}\lesssim 90$~GeV, while by 10-20\% for
100~GeV~$\lesssim m_{H^{\pm\pm}}\lesssim 160$~GeV. 
The ratio of all hadronic decay mode is also enhanced for
$m_{H^{\pm\pm}}<2m_W$ by 5\%, while the ratio of $\ell^\pm\nu jj$ and
d.f.\ $\ell^\pm\ell^\pm\nu\nu$ decay modes is suppressed by 10\%
and 5\%, respectively. 
Therefore, for $m_{H^{\pm\pm}}<2m_W$, the interference term can have
sizable and constructive contribution to the decay rate, and
consequently the s.f.\ $\ell^\pm\ell^\pm\nu\nu$ decay becomes
relatively important. 

The tree level formula for the dilepton decay rate of $H^{\pm\pm}$
is given by 
\begin{align}
 \Gamma(H^{\pm\pm}\to \ell_i^\pm \ell_j^\pm) &= 
\frac{S_{ij}}{8\pi v_\Delta^2} |(m_\nu)_{ij}|^2 m_{H^{\pm\pm}},
 \label{Eq:dilepton}
\end{align}
where $S_{ij}=1~(1/2)$ for $i\neq j$~($i = j$). 

For the cascade decay, taking into account the off-shellness of the W
boson, the tree-level formula is given by 
\begin{align}
 \Gamma(H^{\pm\pm}\to H^\pm W^{\pm(*)}) &=
\frac{9g^4\cos^2\beta}{128\pi^3}m_{H^{\pm\pm}} 
 G\left(\frac{m_{H^{\pm}}^2}{m_{H^{\pm\pm}}^2},
 \frac{m_W^2}{m_{H^{\pm\pm}}^2}\right),\label{Eq:cascade}
\end{align}
where the phase space functions are defined as 
\begin{align}
G(x,y)&=\frac{1}{12y}
 \Bigg\{2\left(x-1\right)^3-9\left(x^2-1\right)y+6\left(x-1\right)y^2
 -3\left[1+\left(x-y\right)^2-2y\right]y\log x\notag\\
 & +6\left(1+x-y\right)y\sqrt{-\lambda(x,y)}
 \Big[\tan^{-1}\Big(\frac{x-y-1}{\sqrt{-\lambda(x,y)}}\Big)+\tan^{-1}
 \Big(\frac{x+y-1}{\sqrt{-\lambda(x,y)}}\Big)\Big]\Bigg\},\\
 \lambda(x,y)&=1+x^2+y^2-2xy-2x-2y. 
\end{align}
We note that the decay rate given in Eq.~(\ref{Eq:cascade}) is valid for
the case with $m_{H^{\pm\pm}}-m_{H^{\pm}}<m_W$. 
In Ref.~\cite{AKKY_full}, it is shown that the mass difference
larger than about 60~GeV is excluded by the electroweak precision data. 
Therefore, the on-shell decay mode of $H^{\pm\pm}\to H^{\pm}W^\pm$ is
disfavored. 

\begin{figure}[t]
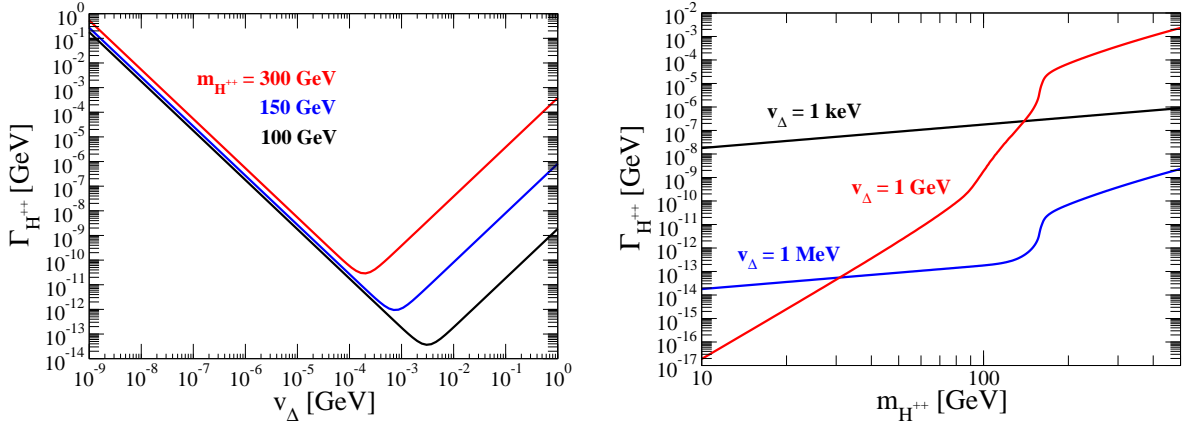

\begin{center}
\includegraphics[width=75mm]{Gamma_Hpp_vdel.eps}\hspace{5mm}
\includegraphics[width=75mm]{Gamma_Hpp_mdch.eps}
\caption{The total width of $H^{\pm\pm}$. 
The left (right) panel shows the $v_\Delta$ ($m_{H^{\pm\pm}}$)
 dependence in the case with $m_{H^{\pm\pm}}=100$, 150 and 300~GeV
 ($v_\Delta=1$~keV, 1~MeV and 1~GeV).
$m_{H^{\pm}}=m_{H^{\pm\pm}}$ is assumed, so that the cascade decay is
 absent. 
}\label{width1}
\end{center}
\end{figure} 

We then evaluate the decay of $H^{\pm\pm}$ for several values of
$v_\Delta$ and $m_{H^{\pm\pm}}$ by taking into account the all three
decay channels; dilepton, diboson and cascade decays. 
For the dilepton decay mode, we take all the elements of the
neutrino mass matrix $(m_\nu)_{ij}$ 0.1~eV. 
In Fig.~\ref{width1}, we show the total decay width of $H^{\pm\pm}$ as
a function of $v_\Delta$ for fixed values of $m_{H^{\pm\pm}}=$100,
150 and 300~GeV (left panel), and as a function of $m_{H^{\pm\pm}}$ for
fixed values of $v_\Delta=$ 1~keV, 1~MeV and 1~GeV (right panel). 
The mass of $H^{\pm}$ is taken to be the same as that of $H^{\pm\pm}$ so
that the cascade decay mode is absent.
As seen in the left panel, the total decay width takes its minimum at
around $v_\Delta=5$~MeV, 1 MeV and 0.2~MeV in the case with
$m_{H^{\pm\pm}}=100$, 150 and 300~GeV, respectively. 
At these minima, the decay rates into the dilepton mode and the diboson
mode are almost the same order. 
In the right panel, in the case where $v_\Delta$ is as small as 1 keV,
the decay width increases linearly with $m_{H^{\pm\pm}}$, because 
the decay width is calculated dominantly from the dilepton decay
rate given in Eq.~(\ref{Eq:dilepton}). 
On the other hand, in the case where $v_\Delta$ is as large as 1 MeV or
1 GeV, the decay rate rapidly increases at around $m_{H^{\pm\pm}}=160$
GeV, because of the threshold of the on-shell W boson pair. 
We note that the decay rate of $10^{-16}$ GeV corresponds to the decay
length of about 1~meter, so that $H^{\pm\pm}$ produced at
colliders would decay inside a detector.

\begin{figure}[t]
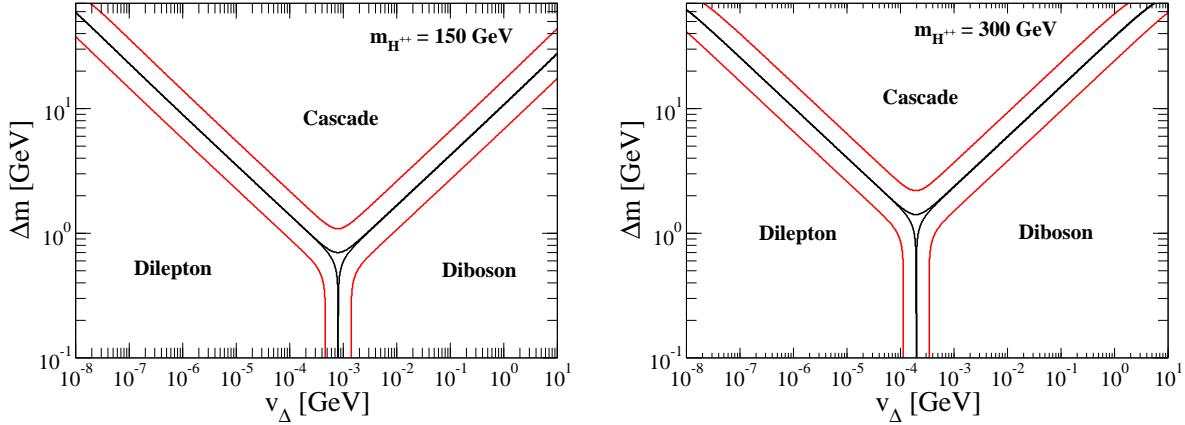

\begin{center}
\includegraphics[width=75mm]{contour_150.eps}\hspace{5mm}
\includegraphics[width=75mm]{contour_300.eps}
\caption{Contour plot for the decay branching ratio of $H^{\pm\pm}$ 
on the $v_\Delta$-$\Delta m~(\equiv m_{H^{\pm\pm}}-m_{H^\pm})$ plane
in the case of $m_{H^{\pm\pm}}=150$ GeV (left panel) and
 $m_{H^{\pm\pm}}=300$ GeV (right panel). 
The parameter regions on the black and red curves give the 50\% and 90\%
 branching ratio for $H^{\pm\pm}$, where the decay mode is indicated
 inside the curves.}
\label{contour}
\end{center}
\end{figure}

In Fig.~\ref{contour}, we show contour plots for the decay branching
ratio of $H^{\pm\pm}$ in the $v_\Delta$-$\Delta m$ plane, where $\Delta
m=m_{H^{\pm\pm}}-m_{H^\pm}$, in the case with $m_{H^{\pm\pm}}=150$ GeV
(left panel) and 300 GeV (right panel). 
In each block bordered by the black (red) contours, the branching ratio
for the decay mode indicated inside the block is greater than 50\%
(90\%).
We can observe that by increasing $v_\Delta$ with fixing $\Delta m$
smaller than about 1 GeV, the main decay mode is replaced from the
dilepton mode to the diboson mode at $v_\Delta\simeq 0.1$-1~MeV. 
The regions where the diboson decay mode dominates are enlarged by
increasing $m_{H^{\pm\pm}}$ from 150~GeV to 300~GeV, due to the cubic
power dependence of the diboson decay rate on $m_{H^{\pm\pm}}$ as
expressed in Eq.~(\ref{Eq:diboson}). 
In the case where $H^{\pm\pm}$ are the lightest among the triplet-like
Higgs bosons, the regions where the cascade decay dominates disappear.

We here comment on the decays of the other triplet-like Higgs
bosons~\cite{Perez,AKY}.
When $v_\Delta$ is smaller than about 1~MeV and $|\Delta m|$ is enough
small, $H^\pm$, $A$ and $H$ mainly decay into $\ell^\pm\nu$, $\nu\nu$
and $\nu\nu$, respectively, similarly to the decay of $H^{\pm\pm}\to
\ell^\pm\ell^\pm$.
When $v_\Delta$ is large; i.e., $v_\Delta \gtrsim 1$~MeV, $H^\pm$
mainly decay into $W^\pm Z$, $hW^\pm$, $\tau^\pm\nu$ and/or $tb$, 
while $A$ mainly decays into $hZ$, $b\bar{b}$, $\tau^+\tau^-$ and/or
$t\bar{t}$. 
The decay of $H$ depends on the mixing angle $\alpha$ in addition to
$v_\Delta$ and $\Delta m$.
As seen in Eq.~(\ref{Eq:mixing_angles}), $\tan2\alpha$ is proportional
to $v_\Delta/v_\phi$, so that small but non-zero $\alpha$ is typically
provided at the same order as $\beta$ and $\beta'$, unless a large
value of $\lambda$ couplings is introduced.
In such a case, the dominant decay mode of $H$ can
be $WW$, $ZZ$, $hh$, $b\bar{b}$, $\tau^+\tau^-$ and/or $t\bar{t}$. 
In the case with non-zero mass difference, a cascade decay $H\to H^\pm
W^{\mp(*)}$ can take place. 

In the following studies, we focus on the same-sign diboson decay
scenario where $\mathcal{B}(H^{\pm\pm}\to W^{\pm(*)}W^{\pm(*)})$ is
assumed to be almost 100\%. 
This scenario can be realized in the case with rather
large $v_\Delta$ with $m_{A/H}\geq m_{H^\pm}\geq m_{H^{\pm\pm}}$ or
$\Delta m\ll 1$~GeV as discussed in this subsection.

\subsection{Production cross sections at the LHC}

The leading production processes of $H^{\pm\pm}$ at the LHC are
\begin{align}
& pp\to H^{++}H^{--} + X, \label{eq:pair} \\
& pp\to H^{\pm\pm}H^{\mp} + X. \label{eq:asso} 
\end{align}
In perturbative QCD, the total cross sections for these processes are
expressed as 
\begin{align}
 &\sigma(pp\to H^{++}H^{--}) = \sum_{q}\int^1_{\tau_0} d\tau
 \frac{d{\mathcal L}_{q\bar{q}}}{d\tau}(\tau,\mu_F)\,
 \hat\sigma_{q\bar{q}\to H^{++}H^{--}}(\tau s),  \label{XS_pair} \\
 &\sigma(pp\to H^{\pm\pm}H^{\mp}) = \sum_{q,q'}\int^1_{\tau_0} d\tau
 \frac{d{\mathcal L}_{q\bar{q}'}}{d\tau}(\tau,\mu_F)\,
 \hat\sigma_{q\bar{q}'\to H^{\pm\pm}H^{\mp}}(\tau s), \label{XS_ass}
\end{align}
where $\tau_0=4m_{H^{\pm\pm}}^2/s$ for Eq.~(\ref{XS_pair}) and 
$\tau_0=(m_{H^{\pm\pm}}+m_{H^\pm})^2/s$ for Eq.~(\ref{XS_ass}). 
$\mu_F$ is the factorization scale.
The partonic cross sections are given at the LO as 
\begin{align}
 & \hat\sigma_{q\bar{q}\to H^{++}H^{--}}(\hat s) =
 \frac{\pi\alpha^2}{9\hat{s}}\left(1-4x_{H^{\pm\pm}}\right)^\frac{3}{2}\left[
 Q_H^2Q_q^2 
 + \frac{(1-x_Z)
 Q_HQ_qV_qV_H
 + \frac{1}{4}(V_q^2+A_q^2)V^2_H}
 {(1-x_Z)^2 
 +  x_Z^2 \Gamma_Z^2/m_Z^2} \right],\\
 & \hat\sigma_{q\bar{q}'\to H^{\pm\pm}H^{\mp}}(\hat s) =
 \frac{\pi\alpha^2\cos^2\beta}{36\hat
 ss_W^4}\left|\frac{1}{1-x_W(1+i\Gamma_W/m_W)}
 \right|^2 \lambda^{3/2}(x_{H^{\pm\pm}},x_{H^\pm}), \label{eq:cs_lhc}
\end{align}
where $x_i = m_i^2/\hat{s}$ ($i=W,~Z,~H^{\pm\pm}$ or $H^\pm$),
$V_q=(T_q^3-2Q_qs_W^2)/(s_Wc_W)$, $A_q=T_q^3/(s_Wc_W)$,
$Q_H=+2$ is the electric charge of $H^{++}$, 
$V_H=(1-2s_W^2)/(s_Wc_W)$, and $s_W=\sin\theta_W$, $c_W=\cos\theta_W$. 
The electric charge and the third component of the isospin for a fermion
$f$ are denoted by $Q_f$ and $T_f^3$, respectively. 
The partonic luminosity functions are defined as
\begin{align}
 & \frac{d{\mathcal L}_{q\bar{q}}}{d\tau}(\tau,\mu_F) = 
 \int^1_0 dx_1 \int^1_0 dx_2 \delta(\tau-x_1x_2)
 \left\{f_q(x_1,\mu_F)f_{\bar{q}}(x_2,\mu_F)
 + f_{\bar{q}}(x_1,\mu_F)f_q(x_2,\mu_F)\right\}, \\
 & \frac{d{\mathcal L}_{q\bar{q}'}}{d\tau}(\tau,\mu_F) = 
 \int^1_0 dx_1 \int^1_0 dx_2 \delta(\tau-x_1x_2)
 \left\{f_q(x_1,\mu_F)f_{\bar{q}'}(x_2,\mu_F)
 + f_{\bar{q}'}(x_1,\mu_F)f_q(x_2,\mu_F)\right\}.
\end{align}

The NLO QCD corrections to the total cross sections are calculated in 
Ref.~\cite{Kfactor}. 
We evaluate the LO and NLO total cross sections for the processes
in Eq.~(\ref{eq:pair}) and Eq.~(\ref{eq:asso}) at the LHC. 
\begin{figure}[t]
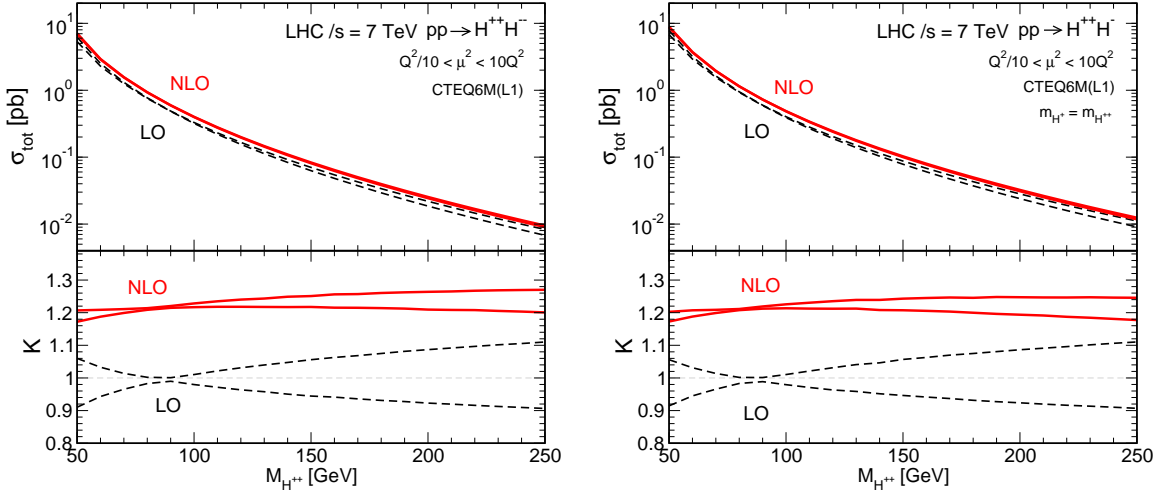

 \begin{center}
  \includegraphics[width=0.45\textwidth]{hpphmm.eps}
  \quad
  \includegraphics[width=0.45\textwidth]{hpphm.eps}
  \caption{Cross sections of $pp\to H^{++}H^{--}$ (left) and $pp\to
  H^{++}H^-$ (right) processes at the LHC with $\sqrt s=7$~TeV evaluated
  at LO and NLO with CTEQ6L1 and CTEQ6M PDFs, respectively. 
  The $K$-factors are also plotted, which are defined as the
  cross sections evaluated at the LO and the NLO with varying the
  scales $\mu^2=\mu^2_R=\mu^2_F$ for $Q^2/10<\mu^2<10Q^2$ divided by the 
  LO cross section evaluated with $\mu^2=Q^2$.
  For the second process, $m_{H^{\pm}}=m_{H^{\pm\pm}}$ is assumed.
  }\label{fig:LHC}
 \end{center}
\end{figure}
In Fig.~\ref{fig:LHC}, we show the cross sections and $K$-factors for
$H^{++}H^{--}$ (left) and $H^{++}H^{-}$ (right) production at the LHC
with $\sqrt{s}=7$~TeV.
We use CTEQ6L1 and CTEQ6M PDFs~\cite{CTEQ} for LO and NLO calculations,
respectively, and vary the factorization scale $\mu_F$ and the
renormalization scale $\mu_R$, where the latter enters at the NLO, to
see uncertainties of the cross section calculations. 
In the top panel, NLO (LO) cross sections are plotted in solid (dashed)
lines as a function of $m_{H^{\pm\pm}}$. 
For each order, two lines are drawn which correspond to the maximum 
and minimal values by varying $\mu=\mu_R=\mu_F$ from $\mu^2=Q^2/10$
to $\mu^2=10Q^2$, where $Q$ is the invariant mass of the final-state
scalar pair. 
Thus, the difference of the two lines indicates the uncertainty of the
calculation by the choice of the scales. 
In the bottom panel, the corresponding $K$-factors are plotted, which
are defined as the ratios of those cross sections to the LO
cross section evaluated with $\mu^2=Q^2$. 
For both processes, the $K$-factors are about 1.2.
The scale uncertainties are typically 5\% (10\%) level for the NLO (LO)
calculation, while these are suppressed accidentally at around
$m_{H^{\pm\pm}}\simeq80$~GeV. 
The uncertainties from PDFs are found to be
about $3\%$ for the lower mass regions but about $10\%$ for the higher
mass regions. 
For reader's convenience, in Appendix~\ref{sec:tab}, we present
tables for the cross sections and their uncertainties for all the
processes in Eq.~(\ref{eq:pair}) and (\ref{eq:asso}) for various values
of $m_{H^{\pm\pm}}$ and various collision energies at the LHC. 

The other $H^{\pm\pm}$ production processes; i.e.,
the vector boson fusion $qQ\to q'Q'H^{\pm\pm}$~\cite{Azuelos:2004dm,%
Godfrey:2010qb,Nomura,Dutta:2014dba}
and the weak boson associated production $q\bar{q}'\to W^{\pm *} \to
H^{\pm\pm}W^\mp$, are induced by the $H^{\pm\pm}W^\mp W^\mp$ coupling
which is proportional to $v_\Delta$ as shown in
Eq.~(\ref{Eq:gauge_coupling}). 
Therefore, these production cross sections are suppressed due to
$v_\Delta/v \ll 1$.

\section{Bound on the mass of $H^{\pm\pm}$}

In this section, we discuss the collider signals for the searches for
$H^{\pm\pm}$ at LEP and the LHC in the diboson decay scenario.
Since there have been no dedicated studies in past experiments, up to
our knowledge, we start to consider the experimental constraint from
relatively small mass regions by using the precise measurement on the Z
boson width at the LEP~I experiment.
Expected signal events for $H^{\pm\pm}$ in the $e^+e^-\to H^{++}H^{--}$
process are studied for the LEP~II energy and luminosity. 
After that, we study the constraint on $H^{\pm\pm}$ in the inclusive 
same-sign dilepton events at the LHC.

\subsection{LEP~I}

The LEP experiment was operated with the electron-positron collision
at the center-of-mass energy on the Z boson mass (LEP~I) and up to about
209~GeV (LEP~II). 
At the LEP~I experiment, the total decay width of the Z boson has been
precisely measured~\cite{LEP_I}.
The measurement can be used to constrain $H^{\pm\pm}$ whose mass is
smaller than a half of $m_Z$ independently of the decay modes of
$H^{\pm\pm}$. 
For $m_{H^{\pm\pm}}$ smaller than a half of $m_Z$, the total decay
width of the Z boson receives a sizable correction from the partial
width for the $Z\to H^{++}H^{--}$ decay as 
\begin{align}
\hspace{-3mm}\Gamma_{Z\to
 H^{++}H^{--}}=\frac{G_Fm_Z^3}{6\pi\sqrt{2}}(1-2s_W^2)^2 
\left(1-\frac{4m_{H^{\pm\pm}}^2}{m_Z^2}\right)^{\frac{3}{2}}. \notag
\end{align}
Using the current experimental data and the SM prediction for the Z
boson width~\cite{PDG}, $\Gamma_Z(\rm exp)=2.4952\pm0.0023$~GeV and
$\Gamma_Z(\rm SM)=2.4960\pm0.0002$~GeV, respectively, we obtain the
lower bound  $m_{H^{\pm\pm}}>42.9$~GeV at the 95\%~CL.

\subsection{LEP~II}

For $m_Z/2<m_{H^{\pm\pm}}<\sqrt{s}/2$, a pair production process of
$H^{\pm\pm}$, $e^+ e^- \to H^{++}H^{--}$, is utilized to search for
$H^{\pm\pm}$ at the LEP~II experiment. 
The total cross section for this process is given by 
\begin{align}
 \sigma_{ee}(s)
 = \frac{\pi\alpha^2}{3s}\left(1-4x_{H^{\pm\pm}}\right)^\frac{3}{2}
 \left[
 Q_H^2Q_e^2 
 + \frac{\left(1-x_Z\right)Q_HQ_eV_eV_H
 + \frac{1}{4}(V_e^2+A_e^2)V^2_H}
 {\left(1-x_Z\right)^2 
 + x_Z^2\Gamma_Z^2/m_Z^2 } \right],
\end{align}
where $V_e=(T_e^3-2Q_es_W^2)/(s_Wc_W)$ and $A_e=T_e^3/(s_Wc_W)$. 
The searches for $H^{\pm\pm}$ in the dilepton decay mode have been
performed at the LEP experiment~\cite{LEP}.
We consider the searches for $H^{\pm\pm}$ in the diboson decay
scenario. 
Through the decays of $H^{\pm\pm}$ into the (off-shell) W bosons, it
subsequently leads to various exotic signals, such as 8-jets, lepton
plus 6-jets plus missing energy, same-sign or opposite-sign dilepton
plus 4-jets plus missing energy, trilepton plus 2-jets plus missing
energy, and tetralepton plus missing energy. 
Produced numbers of events for these signals are estimated to be 
\begin{align}
 N(8\text{-jets}) &= \sigma_{ee}\cdot {\mathcal B}(H^{\pm\pm}\to
 jjjj)^2\cdot {\textstyle \int\mathcal{L}dt},\\
 N(\ell^\pm E_T\hspace{-4.5mm}/\hspace{3mm}+6\text{-jets}) &=
 \sigma_{ee}\cdot 2{\mathcal B}(H^{\pm\pm}\to jjjj){\mathcal 
 B}(H^{\pm\pm}\to \ell\nu jj)\cdot {\textstyle \int\mathcal{L}dt},\\
 N(\ell^\pm\ell^\pm E_T\hspace{-4.5mm}/\hspace{3mm} + 4\text{-jets}) &=
 \sigma_{ee}\cdot 2{\mathcal B}(H^{\pm\pm}\to \ell\ell\nu\nu){\mathcal
 B}(H^{\pm\pm}\to jjjj)\cdot {\textstyle
 \int\mathcal{L}dt},\label{dilepton}\\ 
 N(\ell^\pm\ell^\mp E_T\hspace{-4.5mm}/\hspace{3mm} + 4\text{-jets}) &=
 \sigma_{ee}\cdot {\mathcal B}(H^{\pm\pm}\to\ell\nu jj)^2\cdot
 {\textstyle \int\mathcal{L}dt},  \\ 
 N(\ell^\pm\ell^\pm\ell^\mp
 E_T\hspace{-4.5mm}/\hspace{3mm}+2\text{-jets}) &=
 \sigma_{ee}\cdot2{\mathcal B}(H^{\pm\pm}\to\ell\ell\nu\nu){\mathcal
 B}(H^{\pm\pm}\to\ell\nu jj)\cdot {\textstyle \int\mathcal{L}dt},
 \label{trilepton}\\ 
 N(\ell^+\ell^+\ell^-\ell^- E_T\hspace{-4.5mm}/\hspace{3mm} ) &=
 \sigma_{ee}\cdot{\mathcal B}(H^{\pm\pm}\to\ell\ell\nu\nu)^2\cdot
 {\textstyle \int\mathcal{L}dt},\label{4lepton}
\end{align}
where $\ell=e, \mu$, but the signals with $\tau$'s are neglected for
simplicity.
\begin{figure}[t]
 \begin{center}
  \includegraphics[width=0.6\textwidth]{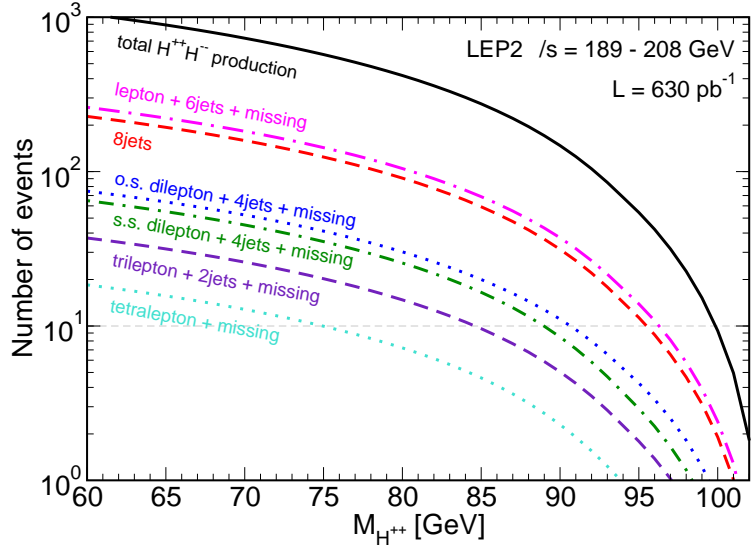}
  \caption{Estimated number of events for various signals in the
  $e^+e^-\to H^{++}H^{--}$ process in the diboson decay scenario at the
  LEP~II experiments as a function of $m_{H^{\pm\pm}}$.
  Total number of events for the $e^+e^-\to H^{++}H^{--}$ production is
  also plotted. 
  The collision energies and the integrated luminosities collected at
  the LEP~II experiments are listed in Table~\ref{tab:lep2}.
  }\label{fig:lep2}
 \end{center}
\end{figure}

In Fig.~\ref{fig:lep2}, we plot the expected number of events for these
signals as a function of $m_{H^{\pm\pm}}$ at the LEP~II experiment.
We calculate the expected number of events by collecting the
cross sections for various collision energies and integrated
luminosities listed in Table~\ref{tab:lep2}~\cite{LEP_II}.
\begin{table}[t]
 \begin{tabular}{rrrrrrrrr}
  \hline
  $\sqrt{s}$~[GeV] &
  188.6 & 191.6 & 195.5 & 199.6 & 201.8 & 204.8 & 206.5 & 208.0 \\
  \hline
  ${\mathcal L}$~[pb$^{-1}$] &
      176.8 & 29.8 & 84.1 & 83.3 & 37.1 & 79.0 & 130.5 & 8.6 \\
  \hline
 \end{tabular}
 \caption{Collision energies $\sqrt{s}$ and integrated luminosities
 ${\mathcal L}$ at the LEP~II experiments~\cite{LEP_II}.}\label{tab:lep2} 
\end{table}
As a reference, the number of event for the total $H^{++}H^{--}$
production is also plotted.

Although the signal of tetralepton plus missing energy
can be compared with the results for the $e^+e^-\to
H^{++}H^{--}\to\ell^+\ell^+\ell^-\ell^-$ search~\cite{LEP} which
requires the four charged-leptons exclusively, 
no substantial bound can been derived in the diboson decay scenario,
because of the suppression of the number of signal events by ${\mathcal
B}(H^{\pm\pm}\to\ell^\pm\ell^\pm\nu\nu)^2$ whose numerical value is
figured in Fig.~\ref{fig:Br4l}.
Up to our knowledge, there have been no dedicated studies on these
signals as direct searches for the same-sign diboson decay of
$H^{\pm\pm}$ at the LEP~II experiment.

For the signals which include same-sign dileptons in
Eq.~(\ref{dilepton}) or trileptons in Eq.~(\ref{trilepton}), we wonder
there can be a chance to find an evidence for $H^{\pm\pm}$ at the LEP~II
experiment.
For example, if we assume that these signals can be discovered if the
expected number of events exceeds ten, $m_{H^{\pm\pm}}\simeq85$~GeV to
90~GeV can be explored at the LEP~II experiment. 
The signal of a lepton plus 6-jets plus missing energy is similar to the
process $e^+e^-\to W^+W^-\to\ell^\pm\nu jj$~\cite{Achard:2004zw}. 
Because the invariant mass of jets is close to $m_W$ for the
latter process, the separation of the two processes seems possible. 
The detection of the 8-jets event should be suffered by background
contribution from QCD events and $W^+W^-$ production in the all hadronic
decays channel~\cite{LEP_II}. 
However, detailed analysis on the event topology variables or shape
variables, such as thrust or acoplanarity, may be used to discriminate
the signal events from the
background~\cite{Abbiendi:1998at,Abbiendi:2004qz}. 
To draw a concrete conclusion, one needs more detailed studies on the
detection efficiencies for these signals, realistic estimation of the
background processes, etc., which are beyond the scope of this paper.
The analyses using the real data at the LEP~II experiments are also
desired. 

Consequently, by using the data as far as we could handle, we obtain the
bound $m_{H^{\pm\pm}}>43$ GeV from the $\Gamma_Z$ measurement at the LEP
experiment, although the bound is quite solid, i.e., independent of the
decay of $H^{\pm\pm}$.

\subsection{Bound from LHC data}

Let us consider the constraint on $m_{H^{\pm\pm}}$ in the diboson decay
scenario by using the current LHC data. 
As explained in Subsection~II B, the main production mode for
$H^{\pm\pm}$ is the pair production $pp\to Z/\gamma^* \to
H^{++}H^{--}$ and the associated production $pp\to W^{\pm} \to
H^{\pm\pm}H^\mp$ at the LHC. 
Among the various final-states in the diboson decay of $H^{\pm\pm}$, the
$\ell^\pm \ell^\pm\nu\nu$ final-state brings the most clean signature at
colliders, since the background contribution can be suppressed for the
signals with same-sign dileptons. 
Thus, we consider that the experimental signatures suited for the
discovery are
\begin{align}
pp &\to H^{++}H^{--}+X\to \ell^\pm\ell^\pm
 E_T\hspace{-4.5mm}/\hspace{3mm}+X, \notag\\ 
pp &\to H^{\pm\pm}H^{\mp}+X \to \ell^\pm\ell^\pm
 E_T\hspace{-4.5mm}/\hspace{3mm}+X, \label{sig1} 
\end{align}
where $\ell^\pm$ denotes $e^\pm$ or $\mu^\pm$. 
The theoretical cross section for the same-sign dilepton signal can be
estimated to be
\begin{align}
 \sigma(\ell^\pm \ell^\pm E_T\hspace{-4.5mm}/\hspace{3mm}+X) &=
 [\sigma(H^{++} H^{--}+X)+\sigma( H^{\pm\pm} H^{\mp}+X)] \times{\mathcal
 B}(H^{\pm\pm}\to\ell^\pm\ell^\pm\nu\nu). \label{sig2}
\end{align}

To obtain the direct bound on $H^{\pm\pm}$, we apply the results
of the same-sign dilepton search reported by the ATLAS
Collaboration~\cite{ATLAS} using the data at the collision energy of 
7~TeV and the integrated luminosity of 4.7~fb$^{-1}$. 
From the data, 95\%~CL upper limits $N_{95}$ for the event number
for the process including the same-sign dilepton have been derived.
In Ref.~\cite{ATLAS}, the limits are separately given for $e^\pm e^\pm$,
$\mu^\pm \mu^\pm$, and $e^\pm\mu^\pm$ channels after imposing several
choices of the cut on the invariant mass $M_{\ell\ell}$ of the same-sign
dilepton.
The 95\%~CL limit for the fiducial cross section
$\sigma^{95}_{\text{fid}}$ is obtained by 
\begin{align}
\sigma^{95}_{\text{fid}}=\frac{N_{95}}{\int \mathcal{L} dt \cdot
 \varepsilon_{\text{fid}}},
\end{align}
where $\varepsilon_{\text{fid}}$ is the efficiency for detecting events
within the detector acceptance, and $\int {\cal L} dt$ is the integrated
luminosity 4.7~fb$^{-1}$. 
The efficiencies are also given in Ref.~\cite{ATLAS} reading 43-65\% for
the $ee$ channel, 55-70\% for the $e\mu$ channel and 59-72\% for the
$\mu\mu$ channel depending on the assumption about momentum
distributions of the charged leptons. 
We find that the data for the $\mu^+\mu^+$ channel with the
invariant-mass cut $M_{\ell\ell}>15$~GeV gives the most severe
constraint on $m_{H^{\pm\pm}}$.
Thus, hereafter, we present a detailed comparison of $\sigma^{95}_{\rm
fid}$ given in Ref.~\cite{ATLAS} with the fiducial cross section
evaluated by ourselves for the $\mu^+\mu^+$ channel with a cut of
$M_{\ell\ell}>15$~GeV.
Theoretical estimation of the fiducial cross section is given by
\begin{align}
 \sigma_{\rm fid} = \sigma_{\rm tot}\cdot{\mathcal B}\cdot\epsilon_{\rm
A}, 
\end{align}
where $\epsilon_{\rm A}$ is the combined efficiency of kinematical
acceptance and kinematical cuts. 
In Table~\ref{tab:lhc}, our estimation for each factor is summarized. 
In the first and second rows, the total cross sections for $pp\to H^{++} 
H^{--}$ and $pp\to H^{++} H^{-}$ processes at the NLO are listed as
a function of $m_{H^{\pm\pm}}$, where $m_{H^\pm}=m_{H^{\pm\pm}}$ is
assumed for the second process. 
We set 5\% uncertainty for the cross sections independently of
$m_{H^{\pm\pm}}$ from the scale uncertainty and the PDF uncertainty.
The branching ratio of $H^{\pm\pm}$ into the same-sign dimuon plus
missing momentum is also listed in the third row in Table~\ref{tab:lhc}.

\begin{table}[t]
 \begin{tabular}{l|rrrrrrrc}
  \hline
  $m_{H^{\pm\pm}}$& 40 & 50 & 60 & 70 & 80 & 90 & 100 &  [GeV] \\
  \hline \hline
  $\sigma^{\rm NLO}_{\rm tot}(pp\to H^{++}H^{--})$ & 120. & 6.95 & 2.90
	      & 1.56 & 0.93 & 0.594 & 0.398 & [pb] \\
  $\sigma^{\rm NLO}_{\rm tot}(pp\to H^{++}H^{-})$ [$m_{H^{\pm}} =
  m_{H^{\pm\pm}}$] & 65. & 8.76 & 3.69 & 1.94 & 1.14 & 0.725 & 0.485 & [pb]
				  \\
  ${\mathcal B}(H^{++}\to\mu^+\mu^+\nu\nu)$ & 2.22 & 2.22 & 2.21 & 2.19
		  & 2.16 & 1.98 & 1.61 & [\%] \\
  $\epsilon_{\rm A}$ ($p_T^\mu>20$~GeV \& $|\eta_\mu|<2.5$) & 0.63 &
	  6.1 & 12. & 17. & 22. & 24. & 23. & [\%] \\
  $\epsilon_{\rm A}$ ($M_{\mu\mu}>15$~GeV) & 78. & 89. & 94. & 96. &
		      98. & 98. & 99. & [\%] \\ \hline
  $\sigma_{\rm fid}(pp\to\mu^{+}\mu^{+}+X)$ [$m_{H^{\pm}} = m_{H^{\pm\pm}}$]
  & 20.2 & 18.9 & 16.4 & 12.5 & 9.6 & 6.1 & 3.2 & [fb]\\
  \hline
 \end{tabular}
 \caption{Table of the total cross sections, branching ratio of
 $H^{\pm\pm}$, and the efficiencies of acceptance and kinematical cuts
 for the $\mu^+\mu^+$ searches at the LHC with 7~TeV for
 $m_{H^{\pm\pm}}=40$~GeV to 100~GeV.
 The resulting fiducial cross section is also listed. }\label{tab:lhc} 
\end{table}

\begin{figure}[t]
 \begin{center}
  \includegraphics[width=\textwidth]{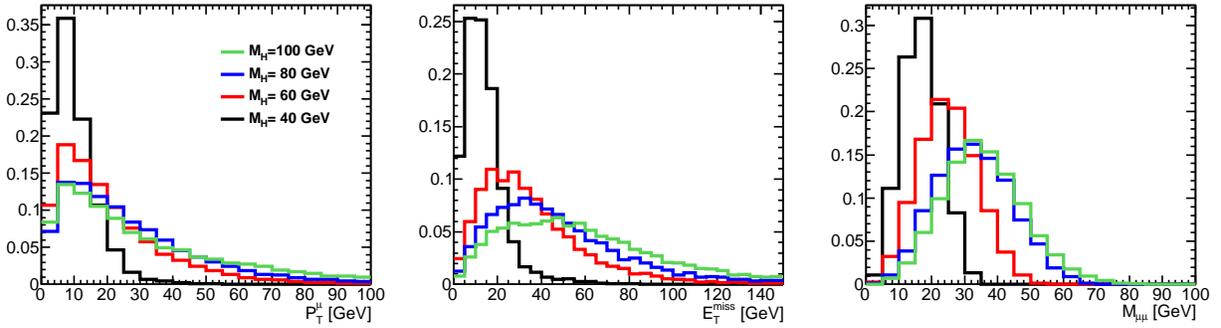}
  \caption{
  Normalized distributions of $p^{}_T$ of muons, missing transverse
  momentum, and the invariant mass of the same-sign dimuon  
  for the inclusive $pp \to (H^{++}\to\mu^+\mu^+\nu\nu)X$ process at
  the LHC with $\sqrt{s}=8$~TeV.
  Distributions are evaluated by using {\tt Madgraph}~\cite{MG5} with
  $m_{H^{\pm\pm}}=40$~GeV, 60~GeV, 80~GeV and 100~GeV.
  }\label{fig:dist}
 \end{center}
\end{figure}

The efficiencies of detector acceptance and kinematical cuts are
separately estimated by using the Monte-Carlo simulation at the parton
level. 
In order to generate the signal events, we use {\tt
MadGraph5}~\cite{MG5} and {\tt CTEQ6L} PDFs~\cite{CTEQ}. 
In Fig.~\ref{fig:dist}, we show the distributions for the signal events
in the transverse momentum of a muon, the missing transverse momentum
and $M_{\mu\mu}$ for $m_{H^{\pm\pm}}=40$~GeV to 100~GeV to check the
shape of the distributions and their mass dependence.
According to Ref.~\cite{ATLAS}, the kinematical cuts by detector
acceptance are taken as 
\begin{align}
&p^\mu_T>20~{\rm GeV},\quad |\eta_\mu|<2.5,\label{basic}
\end{align}
where $p_T^\mu$ and $\eta^\mu$ represent the transverse momentum
and the pseudorapidity of a muon, respectively.
In addition, a cut on $M_{\mu\mu}>15$~GeV is applied.
In the fourth and fifth rows in Table~\ref{tab:lhc}, the efficiencies by
the kinematical acceptance for muons and that by the kinematical cut on
the dimuon invariant-mass are listed, respectively. 
Because only the muon momenta are measured, our parton-level analysis is 
expected to be a good approximation to the realistic detector-level
observation. 
Finally, in the last row, we list the fiducial cross section
as a function of $m_{H^{\pm\pm}}$, calculated by using the numbers in
the upper rows. 
The fiducial cross section takes its maximum value at around
$m_{H^{\pm\pm}}=40$~GeV, because that for lower mass is significantly
reduced by the acceptance cut. 

\begin{figure}[t]
 \begin{center}
  \includegraphics[width=0.6\textwidth]{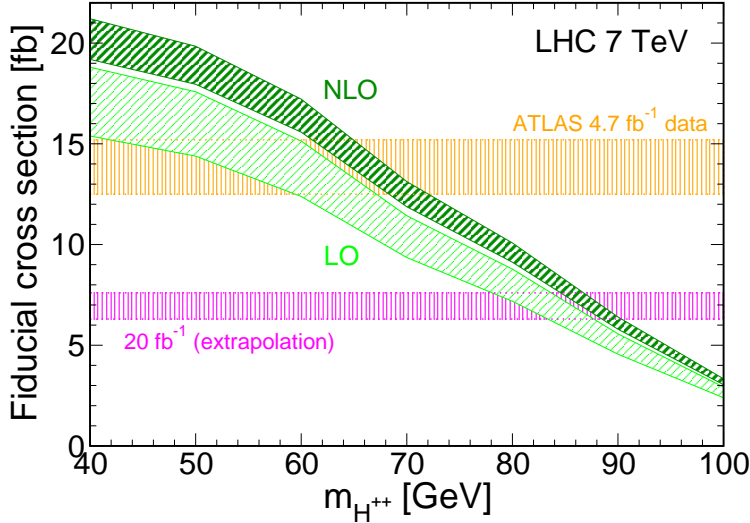}
  \caption{The fiducial cross section for the $\mu^+\mu^+$ events
  at the LHC with $\sqrt{s}=7$~TeV as a function of $m_{H^{\pm\pm}}$. 
  Dark-green and light-green bands show the estimated fiducial
  cross-sections at the NLO and LO, respectively.
  The widths of the band come from the 5\% (10\%) uncertainty for the
  production cross-sections at the NLO (LO).
  The horizontal thick (thin) band shows the (expected) 95\%~CL
  upper limit from the data with the integrated luminosity of
  4.7~fb$^{-1}$ (20~fb$^{-1}$).} \label{fig2}
 \end{center}
\end{figure}

In Fig.~\ref{fig2}, the fiducial cross section for the $\mu^+\mu^+$
events is plotted as a function of $m_{H^{\pm\pm}}$ by a dark-green
band, where its width indicates 5\% uncertainty for the total cross
section at the NLO. 
For the comparison, the LO results previously obtained in
Ref.~\cite{KYY} is also shown in the light green band where 10\%
theory uncertainty is taken into account. 
An orange shaded band gives the 95\%~CL upper limit for the fiducial
cross section obtained in Ref.~\cite{ATLAS} using 4.7~fb$^{-1}$ data. 
The width of the data band comes from the uncertainty of
$\varepsilon_{\text{fid}}$ for the $\mu\mu$ system between 59\% and
72\%~\cite{ATLAS}.
Taking a conservative examination, $H^{\pm\pm}$ is excluded for
$m_{H^{\pm\pm}}\lesssim60$ to 68~GeV in the diboson decay scenario, 
depending on the value of $\epsilon_{\rm fid}$ in Ref.~\cite{ATLAS}. 
We emphasize again that this is the first verification by using the
collider data on the searches for $H^{\pm\pm}$ in the diboson scenario. 
We find that a stronger mass bound is obtained by using the LHC data
more than the bound obtained via the $\Gamma_Z$ measurement at the LEP
experiment. 
The red shaded band is drawn by extrapolating the ATLAS results to
those for 20~fb$^{-1}$ by assuming that the upper limit of the cross
section $\sigma^{95}_{\text{fid}}$ becomes small by a factor 2.
By comparing the extrapolated band with the theoretical cross section,
we obtain that the regions of $m_{H^{\pm\pm}}\lesssim 85$ to 90~GeV
can be surveyed by using the existing LHC data with 20~fb$^{-1}$. 
We note that the difference of the signal cross-sections from 7~TeV to
8~TeV is not taken into account for this extrapolation, since we don't
know how the background cross sections scale at the same time.

We remark that the lower bound obtained in this analysis can be improved
by taking into account the followings:
(i) The other source of extra $H^{\pm\pm}$ from the decay of
$H^{\pm}$ is not considered here, for simplicity, since the decay rate
of $H^{\pm}\to H^{\pm\pm}W^{\mp}$ depends on the other
parameters in the model~\cite{AKY}. 
To count the $H^{\pm\pm}$ production from the decay of $H^{\pm}$, all
the processes of $H^\pm$ production have to be also taken into
account~\cite{AKY,ChunSharma,Sugiyama,Sugiyama2}; such as $pp\to
H^+H^-$, $pp\to H^\pm H$ and $pp\to H^\pm A$. 
(ii) although we have studied the same-sign dileptons only from the
decay of $H^{\pm\pm}$, they can also appear in the decay of $H^{\pm}$;
e.g., $H^{\pm}\to W^{\pm}Z\to\ell^{\pm}\ell^+\ell^-\nu$~\cite{AKY}.
However, the decay of $H^\pm$ also strongly depends on the mass
difference among the triplet Higgs bosons, $\Delta
m=m_{H^\pm}-m_{H^{\pm\pm}}$, so that we here neglect these contributions
as a conservative assumption.
(iii) It is studied that a requirement of relatively hard jets in
addition to the same-sign dilepton in the event can enhance the
significance for discovering $H^{\pm\pm}$~\cite{Nomura,Kang:2014jia}.
In the future LHC run with $\sqrt{s}=13$ to 14~TeV, the mass bound can
be further improved by such an optimized analysis.

Finally, we comment on further observations after the discovery of
$H^{\pm\pm}$, such as the determination of its properties and searches
for the other triplet-like Higgs bosons. 
If $H^{\pm\pm}$ are discovered by the $\ell^\pm\ell^\pm\nu\nu$ events,
the observations of the other signals which come from the hadronic
decays of the (off-shell) W bosons are important to indeed conclude the
diboson decay of $H^{\pm\pm}$. 
The electric charge of $H^{\pm\pm}$ shall be determined from the charges
of the same-sign dilepton as discovery signals in either
the dilepton decay scenario or the diboson decay scenario.
In the dilepton decay scenario, the mass of $H^{\pm\pm}$ can be easily
determined by the sharp peak in the invariant-mass distribution of the
same-sign dilepton.
Angular distributions of leptons discriminate the spin of $H^{\pm\pm}$.
In addition, if $H^{\pm\pm}\to\tau^\pm\tau^\pm$ decays are available,
the spin of $H^{\pm\pm}$ can be directly observed by using the spin
correlation of the two reconstructed $\tau$
leptons~\cite{Bullock:1992yt}. 
On the other hand, in the diboson decay scenario, the determination of
the mass of $H^{\pm\pm}$ is not straightforward, but still possible at
hadron colliders by using the endpoint behavior of the transverse mass
distribution constructed from the dilepton momenta and the missing 
transverse momentum~\cite{AKY}. 
Moreover, the method using the lepton energy
distribution~\cite{Kawabata} may be also applicable since clean signal
events can be extracted in the sense of the kinematical cuts for leptons
and the SM background contributions. 
Observation of the spin of $H^{\pm\pm}$ may be performed in the same
method as the observation of the spin-0 nature of the SM-like Higgs 
bosons at the LHC~\cite{Higgs_property}.
The searches for the other triplet-like Higgs bosons will be performed
at the future LHC run and also at the future lepton colliders, such as
the International Linear Collider (ILC)~\cite{Baer:2013cma} and Compact
Linear Collider (CLIC)~\cite{Accomando:2004sz,Linssen:2012hp}.
Since the heavier triplet-like bosons would decay in the cascade-type,
the searches at the LHC may be difficult and there can be an advantage
for the searches at the future lepton
colliders~\cite{Yagyu:2014aaa}.
Searches for the triplet-like Higgs bosons at photon colliders are also
discussed in Refs.~\cite{Cascade,Cao:2014nta}.

We close this section with comments on the diboson decay of $H^{\pm\pm}$
in the other SU(2)$_{L}$ multiplet.
In general, an SU(2)$_L$ scalar multiplet $\varphi$ which contains both
the doubly-charged $\varphi^{\pm\pm}$ and neutral $\varphi^0$
components can have the $\varphi^{\pm\pm}W^\mp W^\mp$ vertex at the tree
level, when $\varphi^0$ acquires a non-zero VEV\footnote{%
Effective $S^{\pm\pm}W^{\mp}W^{\mp}$ vertex for the singlet
scalar boson $S^{\pm\pm}$ is recently discussed in
Ref.~\cite{King:2014uha}.}.
Although the bound obtained in this paper is limited for the triplet
scalar, the same searches can be applied to them by adopting the
appropriate production cross section and branching ratio.

\section{Conclusion}
We have investigated the collider phenomenology of $H^{\pm\pm}$ in the
HTM, focusing on the scenario where $H^{\pm\pm}$ mainly decay into the
same-sign diboson.
Such a diboson decay scenario can be realized in the case with $v_\Delta
>0.1$-1~MeV and $\Delta m\lesssim1$~GeV as shown in Fig.~\ref{contour}. 
We have shown that the decay branching ratio for the $H^{\pm\pm}\to
\ell^\pm\ell^\pm \nu\nu$ decay with a s.f.\ dilepton is enhanced by up
to 80\% for $m_{H^{\pm\pm}}<2m_W$ due to the interference effect.
Total production cross sections of $pp\to H^{++}H^{--}$ and $pp\to
H^{\pm\pm}H^{\mp}$ are calculated up to the NLO in QCD. 
The predicted cross sections are enhanced by about 20\% from those at
the LO.
These arguments are found to be important to search for $H^{\pm\pm}$ in
the diboson decay scenario in relatively small mass regions.

Since there have been no dedicated studies for the search for
$H^{\pm\pm}$ in the diboson scenario, we have discussed the constraints
at the past collider experiments. 
At the LEP experiment, by comparing the total decay width of the Z
boson with the partial decay rate of $Z\to H^{++}H^{--}$, we have found
that $m_{H^{\pm\pm}}<42.9$~GeV is excluded at the 95\% CL. 
We also have calculated the number of events for various final states
deduced from $e^+e^-\to H^{++}H^{--}$ at the LEP~II. 
Although the signal with tetra leptons plus missing momentum cannot be
used to derive a constraint on $m_{H^{\pm\pm}}$, due to the reduction of
the signal cross section by a square of the branching fraction of
$H^{\pm\pm}\to\ell^{\pm}\ell^{\pm}\nu\nu$ decay, the other signals which
include a same-sign dilepton can be useful for the search for
$H^{\pm\pm}$ in the diboson decay scenario.
We have finished our discussion with emphasizing a need of dedicated
analysis for these signals by using the data from the LEP~II experiment. 

We then have discussed the searches for $H^{\pm\pm}$ in the diboson
decay scenario at the LHC, and also have discussed the bound on the mass
of $H^{\pm\pm}$ from the current data. 
In order to find an evidence of $H^{\pm\pm}$ in relatively lower mass
regions, we treat the theoretical framework for the inclusive same-sign
dilepton signal, which consists of total cross sections for $H^{\pm\pm}$
production at the NLO, the decay branching ratio into the same-flavour
dilepton decay with interference effects, and efficiencies for detector
acceptance and the kinematical cuts.
By combining them, we have evaluated the theoretical prediction for the
fiducial cross section for the same-sign dimuon events. 
By comparing it with the upper limit reported by ATLAS Collaboration
using the 4.7~fb$^{-1}$ data at the 7~TeV run, we find that the lower 
limit of $m_{H^{\pm\pm}}$ in the diboson decay scenario is revised to
60-68~GeV depending on the estimation of the signal efficiency in
the search. 
We have estimated by naive extrapolation that the limit can be extended
up to 85-90~GeV, if full analysis with the available 20~fb$^{-1}$
data set at the 8~TeV run is performed.

Our analysis shows that relatively light $H^{\pm\pm}$ with
$m_{H^{\pm\pm}}\simeq100$~GeV are still allowed if $H^{\pm\pm}$ 
dominantly decay into a (off-shell) dibosons.
In the near future at the LHC run with 13 to 14~TeV, the searches for
$H^{\pm\pm}$ in the diboson decay scenario will be performed to push the
limit toward a few hundreds GeV~\cite{Nomura,Kang:2014jia}. 
At the future lepton colliders, such as the ILC~\cite{Baer:2013cma} and 
CLIC~\cite{Accomando:2004sz,Linssen:2012hp}, we also have a chance to
study the properties of not only $H^{\pm\pm}$ but also the other
triplet-like Higgs bosons~\cite{Yagyu:2014aaa} as long as the masses of
them are within the reach of these colliders.

\section*{Acknowledgments}

We thank Koji Terashi for useful discussions.
This work was supported in part by Grant-in-Aid for Scientific Research,
Nos.\ 22244031, 23104006 and 24340046, JSPS, No.\ 25$\cdot$10031, and
the National Science Council of R.O.C.\ under Grant No.\
NSC-101-2811-M-008-014.

\appendix

\section{$H^{\pm\pm}$ production cross sections at the LHC}\label{sec:tab}

In this appendix, we present the cross sections for $H^{\pm\pm}$
production at the LHC.
We consider the three processes, $pp\to H^{++}H^{--}$, $H^{++}H^{-}$ and
$H^{+}H^{--}$ at the LHC with $\sqrt{s}=7$, 8, 13 and 14~TeV.
We evaluate the total cross sections at the NLO in QCD~\cite{Kfactor}
with {\tt CT10} PDFs~\cite{Lai:2010vv}, and also their uncertainties 
by taking into account the scale ambiguity and the PDF uncertainty. 
For the latter two processes, the mass of $H^{\pm}$ is taken to be
$m_{H^{\pm}}=m_{H^{\pm\pm}}$ for simplicity.

The scale ambiguity is estimated by seeking the maximum and minimum
cross sections by varying the factorization and renormalization scales
$\mu=\mu_F=\mu_R$ in the range $Q^2/10<\mu<10Q^2$ where $Q$ is the
invariant mass of the final-state scalar pair. 
The PDF uncertainties are calculated according to the Hessian method
with 26 eigenvector set provided in Ref.~\cite{Lai:2010vv}.

In Table~\ref{tab:CS7}, the total cross section at the NLO with $\mu=Q$,
its uncertainties from the scale choice and the PDF, in 
order, are presented for the three processes for various values of
$m_{H^{\pm\pm}}$ at the LHC with $\sqrt{s}=7$~TeV. 
The same results but for $\sqrt{s}=8$~TeV, 13~TeV and 14~TeV are also
presented in Table~\ref{tab:CS8}, \ref{tab:CS13} and \ref{tab:CS14},
respectively.

\begin{table}[h]
 \begin{tabular}{|c||lc|lc|lc|}
 \hline
 \multicolumn{7}{|c|}{LHC $\sqrt{s}=7$~TeV, NLO with {\tt CT10} PDFs}\\[1mm]
 \hline
 Mass~[GeV] & \multicolumn{2}{|c|}{$\sigma(H^{++}H^{--})$~[pb]} &
 \multicolumn{2}{|c|}{$\sigma(H^{++}H^{-})$~[pb]} &
 \multicolumn{2}{|c|}{$\sigma(H^{+}H^{--})$~[pb]} \\[1mm]
 \hline\hline
  50 & $7.19\times10^{ 0}$ & $^{+1.7\%}_{-0.8\%}{}^{+ 3.2\%}_{- 3.5\%}$
     & $9.01\times10^{ 0}$ & $^{+1.6\%}_{-0.7\%}{}^{+ 3.8\%}_{- 4.4\%}$
     & $5.60\times10^{ 0}$ & $^{+1.8\%}_{-0.7\%}{}^{+ 3.7\%}_{- 4.2\%}$\\[1mm]
  60 & $2.98\times10^{ 0}$ & $^{+1.1\%}_{-0.3\%}{}^{+ 3.4\%}_{- 3.6\%}$
     & $3.78\times10^{ 0}$ & $^{+1.0\%}_{-0.3\%}{}^{+ 3.9\%}_{- 4.5\%}$
     & $2.26\times10^{ 0}$ & $^{+1.2\%}_{-0.4\%}{}^{+ 3.9\%}_{- 4.3\%}$\\[1mm]
  70 & $1.60\times10^{ 0}$ & $^{+0.6\%}_{-0.1\%}{}^{+ 3.4\%}_{- 3.8\%}$
     & $1.98\times10^{ 0}$ & $^{+0.5\%}_{-0.1\%}{}^{+ 4.0\%}_{- 4.7\%}$
     & $1.14\times10^{ 0}$ & $^{+0.7\%}_{-0.1\%}{}^{+ 4.0\%}_{- 4.4\%}$\\[1mm]
  80 & $9.51\times10^{-1}$ & $^{+0.4\%}_{-0.0\%}{}^{+ 3.4\%}_{- 4.0\%}$
     & $1.16\times10^{ 0}$ & $^{+0.4\%}_{-0.0\%}{}^{+ 4.1\%}_{- 4.9\%}$
     & $6.50\times10^{-1}$ & $^{+0.3\%}_{-0.0\%}{}^{+ 4.2\%}_{- 4.7\%}$\\[1mm]
  90 & $6.07\times10^{-1}$ & $^{+0.5\%}_{-0.1\%}{}^{+ 3.5\%}_{- 4.2\%}$
     & $7.37\times10^{-1}$ & $^{+0.5\%}_{-0.2\%}{}^{+ 4.1\%}_{- 5.1\%}$
     & $4.00\times10^{-1}$ & $^{+0.5\%}_{-0.1\%}{}^{+ 4.4\%}_{- 4.9\%}$\\[1mm]
 100 & $4.06\times10^{-1}$ & $^{+0.7\%}_{-0.4\%}{}^{+ 3.7\%}_{- 4.3\%}$
     & $4.92\times10^{-1}$ & $^{+0.7\%}_{-0.5\%}{}^{+ 4.2\%}_{- 5.3\%}$
     & $2.60\times10^{-1}$ & $^{+0.8\%}_{-0.4\%}{}^{+ 4.5\%}_{- 5.2\%}$\\[1mm]
 120 & $2.01\times10^{-1}$ & $^{+1.0\%}_{-1.0\%}{}^{+ 3.9\%}_{- 4.7\%}$
     & $2.44\times10^{-1}$ & $^{+1.1\%}_{-1.1\%}{}^{+ 4.4\%}_{- 5.8\%}$
     & $1.23\times10^{-1}$ & $^{+1.1\%}_{-1.0\%}{}^{+ 5.0\%}_{- 5.7\%}$\\[1mm]
 140 & $1.10\times10^{-1}$ & $^{+1.3\%}_{-1.5\%}{}^{+ 4.1\%}_{- 5.0\%}$
     & $1.34\times10^{-1}$ & $^{+1.3\%}_{-1.5\%}{}^{+ 4.6\%}_{- 6.3\%}$
     & $6.45\times10^{-2}$ & $^{+1.4\%}_{-1.5\%}{}^{+ 5.3\%}_{- 6.2\%}$\\[1mm]
 160 & $6.41\times10^{-2}$ & $^{+1.5\%}_{-1.9\%}{}^{+ 4.3\%}_{- 5.4\%}$
     & $7.91\times10^{-2}$ & $^{+1.6\%}_{-1.9\%}{}^{+ 5.0\%}_{- 6.7\%}$
     & $3.66\times10^{-2}$ & $^{+1.6\%}_{-2.0\%}{}^{+ 5.7\%}_{- 6.6\%}$\\[1mm]
 180 & $3.94\times10^{-2}$ & $^{+1.8\%}_{-2.3\%}{}^{+ 4.6\%}_{- 5.8\%}$
     & $4.91\times10^{-2}$ & $^{+1.8\%}_{-2.3\%}{}^{+ 5.2\%}_{- 7.2\%}$
     & $2.19\times10^{-2}$ & $^{+1.9\%}_{-2.3\%}{}^{+ 6.1\%}_{- 7.1\%}$\\[1mm]
 200 & $2.52\times10^{-2}$ & $^{+2.0\%}_{-2.6\%}{}^{+ 4.8\%}_{- 6.1\%}$
     & $3.18\times10^{-2}$ & $^{+2.0\%}_{-2.7\%}{}^{+ 5.5\%}_{- 7.7\%}$
     & $1.37\times10^{-2}$ & $^{+2.1\%}_{-2.7\%}{}^{+ 6.5\%}_{- 7.5\%}$\\[1mm]
 250 & $9.35\times10^{-3}$ & $^{+2.5\%}_{-3.4\%}{}^{+ 5.6\%}_{- 6.9\%}$
     & $1.21\times10^{-2}$ & $^{+2.5\%}_{-3.5\%}{}^{+ 6.5\%}_{- 8.9\%}$
     & $4.84\times10^{-3}$ & $^{+2.6\%}_{-3.5\%}{}^{+ 7.7\%}_{- 8.5\%}$\\[1mm]
 300 & $3.95\times10^{-3}$ & $^{+3.0\%}_{-4.1\%}{}^{+ 6.3\%}_{- 7.7\%}$
     & $5.23\times10^{-3}$ & $^{+3.0\%}_{-4.1\%}{}^{+ 7.6\%}_{-10.3\%}$
     & $1.96\times10^{-3}$ & $^{+3.1\%}_{-4.2\%}{}^{+ 8.8\%}_{- 9.6\%}$\\[1mm]
 350 & $1.82\times10^{-3}$ & $^{+3.5\%}_{-4.7\%}{}^{+ 7.1\%}_{- 8.6\%}$
     & $2.45\times10^{-3}$ & $^{+3.5\%}_{-4.8\%}{}^{+ 9.1\%}_{-11.5\%}$
     & $8.74\times10^{-4}$ & $^{+3.6\%}_{-4.8\%}{}^{+10.2\%}_{-10.7\%}$\\[1mm]
 400 & $8.91\times10^{-4}$ & $^{+3.9\%}_{-5.2\%}{}^{+ 8.1\%}_{- 9.4\%}$
     & $1.22\times10^{-3}$ & $^{+4.0\%}_{-5.4\%}{}^{+10.7\%}_{-13.0\%}$
     & $4.16\times10^{-4}$ & $^{+4.0\%}_{-5.3\%}{}^{+11.9\%}_{-12.0\%}$\\[1mm]
 \hline
 \end{tabular}
 \caption{Total cross sections at the NLO with {\tt CT10} PDFs for
 $H^{\pm\pm}$ production processes at the LHC with $\sqrt{s}=7$~TeV.
 The errors show the uncertainty by varying the scale $\mu=\mu_R=\mu_F$
 as $Q^2/10<\mu^2<10Q^2$ and the uncertainty in the 
 PDFs, in order.
 For the second and third processes, $m_{H^\pm}=m_{H^{\pm\pm}}$ is
 assumed. 
 }\label{tab:CS7} 
\end{table}
\begin{table}[h]
 \centering
 \begin{tabular}{|c||lc|lc|lc|}
 \hline
 \multicolumn{7}{|c|}{LHC $\sqrt{s}=8$~TeV, NLO with {\tt CT10} PDFs}\\[1mm]
 \hline
 Mass~[GeV] & \multicolumn{2}{|c|}{$\sigma(H^{++}H^{--})$~[pb]} &
 \multicolumn{2}{|c|}{$\sigma(H^{++}H^{-})$~[pb]} &
 \multicolumn{2}{|c|}{$\sigma(H^{+}H^{--})$~[pb]} \\[1mm]
 \hline\hline
  50 & $8.52\times10^{ 0}$ & $^{+2.0\%}_{-1.1\%}{}^{+ 3.1\%}_{- 3.5\%}$
     & $1.06\times10^{ 1}$ & $^{+2.0\%}_{-1.0\%}{}^{+ 3.7\%}_{- 4.3\%}$
     & $6.71\times10^{ 0}$ & $^{+2.2\%}_{-1.0\%}{}^{+ 3.5\%}_{- 4.1\%}$\\[1mm]
  60 & $3.57\times10^{ 0}$ & $^{+1.4\%}_{-0.6\%}{}^{+ 3.3\%}_{- 3.4\%}$
     & $4.47\times10^{ 0}$ & $^{+1.3\%}_{-0.5\%}{}^{+ 3.8\%}_{- 4.4\%}$
     & $2.73\times10^{ 0}$ & $^{+1.5\%}_{-0.5\%}{}^{+ 3.7\%}_{- 4.1\%}$\\[1mm]
  70 & $1.93\times10^{ 0}$ & $^{+1.0\%}_{-0.2\%}{}^{+ 3.3\%}_{- 3.6\%}$
     & $2.36\times10^{ 0}$ & $^{+0.9\%}_{-0.2\%}{}^{+ 3.9\%}_{- 4.5\%}$
     & $1.40\times10^{ 0}$ & $^{+1.0\%}_{-0.3\%}{}^{+ 3.8\%}_{- 4.3\%}$\\[1mm]
  80 & $1.16\times10^{ 0}$ & $^{+0.6\%}_{-0.1\%}{}^{+ 3.4\%}_{- 3.7\%}$
     & $1.40\times10^{ 0}$ & $^{+0.5\%}_{-0.1\%}{}^{+ 3.9\%}_{- 4.7\%}$
     & $8.03\times10^{-1}$ & $^{+0.6\%}_{-0.1\%}{}^{+ 3.9\%}_{- 4.5\%}$\\[1mm]
  90 & $7.44\times10^{-1}$ & $^{+0.6\%}_{-0.0\%}{}^{+ 3.4\%}_{- 3.9\%}$
     & $8.91\times10^{-1}$ & $^{+0.3\%}_{-0.0\%}{}^{+ 4.0\%}_{- 4.8\%}$
     & $4.98\times10^{-1}$ & $^{+0.3\%}_{-0.0\%}{}^{+ 4.1\%}_{- 4.7\%}$\\[1mm]
 100 & $5.01\times10^{-1}$ & $^{+0.5\%}_{-0.1\%}{}^{+ 3.5\%}_{- 4.0\%}$
     & $5.98\times10^{-1}$ & $^{+0.5\%}_{-0.2\%}{}^{+ 4.0\%}_{- 5.0\%}$
     & $3.26\times10^{-1}$ & $^{+0.5\%}_{-0.1\%}{}^{+ 4.3\%}_{- 4.9\%}$\\[1mm]
 120 & $2.52\times10^{-1}$ & $^{+0.8\%}_{-0.6\%}{}^{+ 3.6\%}_{- 4.5\%}$
     & $3.01\times10^{-1}$ & $^{+0.8\%}_{-0.7\%}{}^{+ 4.2\%}_{- 5.5\%}$
     & $1.57\times10^{-1}$ & $^{+0.8\%}_{-0.6\%}{}^{+ 4.6\%}_{- 5.3\%}$\\[1mm]
 140 & $1.39\times10^{-1}$ & $^{+1.1\%}_{-1.0\%}{}^{+ 3.8\%}_{- 4.8\%}$
     & $1.68\times10^{-1}$ & $^{+1.1\%}_{-1.2\%}{}^{+ 4.4\%}_{- 5.8\%}$
     & $8.37\times10^{-2}$ & $^{+1.1\%}_{-1.1\%}{}^{+ 4.9\%}_{- 5.7\%}$\\[1mm]
 160 & $8.25\times10^{-2}$ & $^{+1.3\%}_{-1.5\%}{}^{+ 4.0\%}_{- 5.0\%}$
     & $1.00\times10^{-1}$ & $^{+1.3\%}_{-1.5\%}{}^{+ 4.7\%}_{- 6.2\%}$
     & $4.81\times10^{-2}$ & $^{+1.4\%}_{-1.5\%}{}^{+ 5.3\%}_{- 6.1\%}$\\[1mm]
 180 & $5.14\times10^{-2}$ & $^{+1.5\%}_{-1.8\%}{}^{+ 4.2\%}_{- 5.4\%}$
     & $6.30\times10^{-2}$ & $^{+1.5\%}_{-1.9\%}{}^{+ 4.8\%}_{- 6.7\%}$
     & $2.92\times10^{-2}$ & $^{+1.6\%}_{-1.9\%}{}^{+ 5.6\%}_{- 6.5\%}$\\[1mm]
 200 & $3.34\times10^{-2}$ & $^{+1.7\%}_{-2.2\%}{}^{+ 4.5\%}_{- 5.7\%}$
     & $4.13\times10^{-2}$ & $^{+1.7\%}_{-2.2\%}{}^{+ 5.1\%}_{- 7.0\%}$
     & $1.85\times10^{-2}$ & $^{+1.8\%}_{-2.2\%}{}^{+ 6.0\%}_{- 7.0\%}$\\[1mm]
 250 & $1.28\times10^{-2}$ & $^{+2.2\%}_{-2.9\%}{}^{+ 5.1\%}_{- 6.4\%}$
     & $1.63\times10^{-2}$ & $^{+2.2\%}_{-2.9\%}{}^{+ 5.9\%}_{- 8.1\%}$
     & $6.81\times10^{-3}$ & $^{+2.3\%}_{-3.0\%}{}^{+ 7.0\%}_{- 7.8\%}$\\[1mm]
 300 & $5.62\times10^{-3}$ & $^{+2.6\%}_{-3.5\%}{}^{+ 5.8\%}_{- 7.1\%}$
     & $7.30\times10^{-3}$ & $^{+2.6\%}_{-3.6\%}{}^{+ 6.8\%}_{- 9.2\%}$
     & $2.87\times10^{-3}$ & $^{+2.7\%}_{-3.6\%}{}^{+ 7.9\%}_{- 8.8\%}$\\[1mm]
 350 & $2.68\times10^{-3}$ & $^{+3.0\%}_{-4.1\%}{}^{+ 6.4\%}_{- 7.8\%}$
     & $3.56\times10^{-3}$ & $^{+3.0\%}_{-4.2\%}{}^{+ 7.8\%}_{-10.4\%}$
     & $1.32\times10^{-3}$ & $^{+3.1\%}_{-4.2\%}{}^{+ 9.1\%}_{- 9.7\%}$\\[1mm]
 400 & $1.37\times10^{-3}$ & $^{+3.4\%}_{-4.6\%}{}^{+ 7.2\%}_{- 8.5\%}$
     & $1.84\times10^{-3}$ & $^{+3.5\%}_{-4.7\%}{}^{+ 9.0\%}_{-11.6\%}$
     & $6.55\times10^{-4}$ & $^{+3.5\%}_{-4.7\%}{}^{+10.2\%}_{-10.7\%}$\\[1mm]
 \hline
 \end{tabular}
\caption{The same as TABLE~\ref{tab:CS7}, but for the LHC with
 $\sqrt{s}=8$~TeV}\label{tab:CS8}
\end{table}
\begin{table}[h]
 \centering
 \begin{tabular}{|c||lc|lc|lc|}
 \hline
 \multicolumn{7}{|c|}{LHC $\sqrt{s}=13$~TeV, NLO with {\tt CT10} PDFs}\\[1mm]
 \hline
 Mass~[GeV] & \multicolumn{2}{|c|}{$\sigma(H^{++}H^{--})$~[pb]} &
 \multicolumn{2}{|c|}{$\sigma(H^{++}H^{-})$~[pb]} &
 \multicolumn{2}{|c|}{$\sigma(H^{+}H^{--})$~[pb]}\\[1mm]
 \hline\hline
  80 & $2.26\times10^{ 0}$ & $^{+1.7\%}_{-0.8\%}{}^{+3.0\%}_{- 3.4\%}$
     & $2.62\times10^{ 0}$ & $^{+1.6\%}_{-0.7\%}{}^{+3.5\%}_{- 4.1\%}$
     & $1.65\times10^{ 0}$ & $^{+1.7\%}_{-0.7\%}{}^{+3.4\%}_{- 3.9\%}$\\[1mm]
  90 & $1.48\times10^{ 0}$ & $^{+1.4\%}_{-0.5\%}{}^{+3.1\%}_{- 3.4\%}$
     & $1.70\times10^{ 0}$ & $^{+1.2\%}_{-0.5\%}{}^{+3.7\%}_{- 4.1\%}$
     & $1.05\times10^{ 0}$ & $^{+1.4\%}_{-0.5\%}{}^{+3.5\%}_{- 3.9\%}$\\[1mm]
 100 & $1.02\times10^{ 0}$ & $^{+1.1\%}_{-0.3\%}{}^{+3.1\%}_{- 3.5\%}$
     & $1.16\times10^{ 0}$ & $^{+1.0\%}_{-0.3\%}{}^{+3.7\%}_{- 4.2\%}$
     & $7.02\times10^{-1}$ & $^{+1.1\%}_{-0.3\%}{}^{+3.6\%}_{- 4.0\%}$\\[1mm]
 120 & $5.33\times10^{-1}$ & $^{+0.6\%}_{-0.1\%}{}^{+3.2\%}_{- 3.6\%}$
     & $6.07\times10^{-1}$ & $^{+0.5\%}_{-0.1\%}{}^{+3.7\%}_{- 4.4\%}$
     & $3.53\times10^{-1}$ & $^{+0.6\%}_{-0.1\%}{}^{+3.8\%}_{- 4.2\%}$\\[1mm]
 140 & $3.07\times10^{-1}$ & $^{+0.4\%}_{-0.0\%}{}^{+3.2\%}_{- 3.8\%}$
     & $3.50\times10^{-1}$ & $^{+0.3\%}_{-0.0\%}{}^{+3.8\%}_{- 4.6\%}$
     & $1.97\times10^{-1}$ & $^{+0.3\%}_{-0.0\%}{}^{+3.9\%}_{- 4.5\%}$\\[1mm]
 160 & $1.89\times10^{-1}$ & $^{+0.5\%}_{-0.2\%}{}^{+3.4\%}_{- 3.9\%}$
     & $2.17\times10^{-1}$ & $^{+0.5\%}_{-0.2\%}{}^{+4.0\%}_{- 4.8\%}$
     & $1.18\times10^{-1}$ & $^{+0.5\%}_{-0.2\%}{}^{+4.2\%}_{- 4.7\%}$\\[1mm]
 180 & $1.22\times10^{-1}$ & $^{+0.7\%}_{-0.4\%}{}^{+3.5\%}_{- 4.1\%}$
     & $1.41\times10^{-1}$ & $^{+0.7\%}_{-0.5\%}{}^{+4.1\%}_{- 5.1\%}$
     & $7.48\times10^{-2}$ & $^{+0.7\%}_{-0.5\%}{}^{+4.3\%}_{- 5.0\%}$\\[1mm]
 200 & $8.23\times10^{-2}$ & $^{+0.8\%}_{-0.7\%}{}^{+3.5\%}_{- 4.3\%}$
     & $9.58\times10^{-2}$ & $^{+0.8\%}_{-0.8\%}{}^{+4.2\%}_{- 5.3\%}$
     & $4.95\times10^{-2}$ & $^{+0.8\%}_{-0.8\%}{}^{+4.6\%}_{- 5.2\%}$\\[1mm]
 250 & $3.48\times10^{-2}$ & $^{+1.2\%}_{-1.4\%}{}^{+3.8\%}_{- 4.9\%}$
     & $4.14\times10^{-2}$ & $^{+1.2\%}_{-1.4\%}{}^{+4.5\%}_{- 6.0\%}$
     & $2.01\times10^{-2}$ & $^{+1.2\%}_{-1.4\%}{}^{+5.1\%}_{- 5.9\%}$\\[1mm]
 300 & $1.68\times10^{-2}$ & $^{+1.5\%}_{-1.9\%}{}^{+4.2\%}_{- 5.4\%}$
     & $2.03\times10^{-2}$ & $^{+1.5\%}_{-1.9\%}{}^{+4.9\%}_{- 6.6\%}$
     & $9.34\times10^{-3}$ & $^{+1.6\%}_{-2.0\%}{}^{+5.6\%}_{- 6.6\%}$\\[1mm]
 350 & $8.82\times10^{-3}$ & $^{+1.8\%}_{-2.3\%}{}^{+4.6\%}_{- 5.8\%}$
     & $1.09\times10^{-2}$ & $^{+1.8\%}_{-2.4\%}{}^{+5.2\%}_{- 7.3\%}$
     & $4.77\times10^{-3}$ & $^{+1.9\%}_{-2.4\%}{}^{+6.2\%}_{- 7.2\%}$\\[1mm]
 400 & $4.95\times10^{-3}$ & $^{+2.0\%}_{-2.8\%}{}^{+5.0\%}_{- 6.2\%}$
     & $6.22\times10^{-3}$ & $^{+2.0\%}_{-2.8\%}{}^{+5.8\%}_{- 7.9\%}$
     & $2.61\times10^{-3}$ & $^{+2.1\%}_{-2.9\%}{}^{+6.8\%}_{- 7.7\%}$\\[1mm]
 450 & $2.92\times10^{-3}$ & $^{+2.3\%}_{-3.1\%}{}^{+5.3\%}_{- 6.8\%}$
     & $3.72\times10^{-3}$ & $^{+2.3\%}_{-3.1\%}{}^{+6.4\%}_{- 8.6\%}$
     & $1.50\times10^{-3}$ & $^{+2.4\%}_{-3.2\%}{}^{+7.5\%}_{- 8.3\%}$\\[1mm]
 500 & $1.79\times10^{-3}$ & $^{+2.5\%}_{-3.4\%}{}^{+5.7\%}_{- 7.2\%}$
     & $2.31\times10^{-3}$ & $^{+2.5\%}_{-3.5\%}{}^{+6.9\%}_{- 9.3\%}$
     & $9.00\times10^{-4}$ & $^{+2.6\%}_{-3.6\%}{}^{+8.1\%}_{- 8.9\%}$\\[1mm]
 550 & $1.13\times10^{-3}$ & $^{+2.7\%}_{-3.8\%}{}^{+6.2\%}_{- 7.6\%}$
     & $1.48\times10^{-3}$ & $^{+2.8\%}_{-3.8\%}{}^{+7.6\%}_{-10.1\%}$
     & $5.58\times10^{-4}$ & $^{+2.9\%}_{-3.9\%}{}^{+8.8\%}_{- 9.4\%}$\\[1mm]
 \hline
 \end{tabular}
\caption{The same as TABLE~\ref{tab:CS7}, but for the LHC with
 $\sqrt{s}=13$~TeV}\label{tab:CS13} 
\end{table}
\begin{table}[h]
 \centering
 \begin{tabular}{|c||lc|lc|lc|}
 \hline
 \multicolumn{7}{|c|}{LHC $\sqrt{s}=14$~TeV, NLO with {\tt CT10} PDFs}\\[1mm]
 \hline
 Mass~[GeV] & \multicolumn{2}{|c|}{$\sigma(H^{++}H^{--})$~[pb]} &
 \multicolumn{2}{|c|}{$\sigma(H^{++}H^{-})$~[pb]} &
 \multicolumn{2}{|c|}{$\sigma(H^{+}H^{--})$~[pb]}\\[1mm]
 \hline\hline
  80 & $2.49\times10^{ 0}$ & $^{+1.8\%}_{-0.9\%}{}^{+3.0\%}_{-3.4\%}$
     & $2.87\times10^{ 0}$ & $^{+1.7\%}_{-0.9\%}{}^{+3.4\%}_{-4.1\%}$
     & $1.83\times10^{ 0}$ & $^{+1.9\%}_{-0.9\%}{}^{+3.3\%}_{-3.8\%}$\\[1mm]
  90 & $1.64\times10^{ 0}$ & $^{+1.5\%}_{-0.7\%}{}^{+3.1\%}_{-3.3\%}$
     & $1.87\times10^{ 0}$ & $^{+1.4\%}_{-0.6\%}{}^{+3.6\%}_{-4.0\%}$
     & $1.17\times10^{ 0}$ & $^{+1.6\%}_{-0.6\%}{}^{+3.5\%}_{-3.8\%}$\\[1mm]
 100 & $1.13\times10^{ 0}$ & $^{+1.2\%}_{-0.5\%}{}^{+3.1\%}_{-3.3\%}$
     & $1.28\times10^{ 0}$ & $^{+1.1\%}_{-0.4\%}{}^{+3.5\%}_{-4.2\%}$
     & $7.84\times10^{-1}$ & $^{+1.3\%}_{-0.5\%}{}^{+3.6\%}_{-3.8\%}$\\[1mm]
 120 & $5.93\times10^{-1}$ & $^{+0.8\%}_{-0.2\%}{}^{+3.1\%}_{-3.5\%}$
     & $6.71\times10^{-1}$ & $^{+0.7\%}_{-0.1\%}{}^{+3.7\%}_{-4.3\%}$
     & $3.96\times10^{-1}$ & $^{+0.8\%}_{-0.2\%}{}^{+3.7\%}_{-4.1\%}$\\[1mm]
 140 & $3.43\times10^{-1}$ & $^{+0.4\%}_{-0.0\%}{}^{+3.2\%}_{-3.7\%}$
     & $3.89\times10^{-1}$ & $^{+0.3\%}_{-0.0\%}{}^{+3.8\%}_{-4.5\%}$
     & $2.22\times10^{-1}$ & $^{+0.3\%}_{-0.0\%}{}^{+3.9\%}_{-4.3\%}$\\[1mm]
 160 & $2.12\times10^{-1}$ & $^{+0.4\%}_{-0.0\%}{}^{+3.3\%}_{-3.8\%}$
     & $2.41\times10^{-1}$ & $^{+0.5\%}_{-0.1\%}{}^{+3.7\%}_{-4.8\%}$
     & $1.34\times10^{-1}$ & $^{+0.4\%}_{-0.0\%}{}^{+4.2\%}_{-4.5\%}$\\[1mm]
 180 & $1.38\times10^{-1}$ & $^{+0.5\%}_{-0.3\%}{}^{+3.4\%}_{-4.0\%}$
     & $1.58\times10^{-1}$ & $^{+0.6\%}_{-0.4\%}{}^{+3.9\%}_{-5.0\%}$
     & $8.53\times10^{-2}$ & $^{+0.5\%}_{-0.3\%}{}^{+4.3\%}_{-4.7\%}$\\[1mm]
 200 & $9.33\times10^{-2}$ & $^{+0.7\%}_{-0.5\%}{}^{+3.4\%}_{-4.2\%}$
     & $1.08\times10^{-1}$ & $^{+0.7\%}_{-0.6\%}{}^{+4.0\%}_{-5.2\%}$
     & $5.67\times10^{-2}$ & $^{+0.7\%}_{-0.6\%}{}^{+4.4\%}_{-5.0\%}$\\[1mm]
 250 & $3.99\times10^{-2}$ & $^{+1.1\%}_{-1.1\%}{}^{+3.8\%}_{-4.6\%}$
     & $4.70\times10^{-2}$ & $^{+1.0\%}_{-1.2\%}{}^{+4.3\%}_{-5.8\%}$
     & $2.33\times10^{-2}$ & $^{+1.1\%}_{-1.2\%}{}^{+4.9\%}_{-5.7\%}$\\[1mm]
 300 & $1.94\times10^{-2}$ & $^{+1.4\%}_{-1.7\%}{}^{+4.0\%}_{-5.1\%}$
     & $2.33\times10^{-2}$ & $^{+1.4\%}_{-1.7\%}{}^{+4.7\%}_{-6.3\%}$
     & $1.10\times10^{-2}$ & $^{+2.5\%}_{-1.7\%}{}^{+5.4\%}_{-6.3\%}$\\[1mm]
 350 & $1.03\times10^{-2}$ & $^{+1.6\%}_{-2.1\%}{}^{+4.3\%}_{-5.6\%}$
     & $1.26\times10^{-2}$ & $^{+1.6\%}_{-2.2\%}{}^{+5.1\%}_{-6.9\%}$
     & $5.66\times10^{-3}$ & $^{+1.7\%}_{-2.2\%}{}^{+5.9\%}_{-6.9\%}$\\[1mm]
 400 & $5.85\times10^{-3}$ & $^{+1.9\%}_{-2.5\%}{}^{+4.7\%}_{-6.0\%}$
     & $7.28\times10^{-3}$ & $^{+1.9\%}_{-2.5\%}{}^{+5.5\%}_{-7.5\%}$
     & $3.13\times10^{-3}$ & $^{+2.0\%}_{-2.6\%}{}^{+6.5\%}_{-7.4\%}$\\[1mm]
 450 & $3.49\times10^{-3}$ & $^{+2.1\%}_{-2.8\%}{}^{+5.1\%}_{-6.4\%}$
     & $4.41\times10^{-3}$ & $^{+2.1\%}_{-2.9\%}{}^{+6.0\%}_{-8.2\%}$
     & $1.82\times10^{-3}$ & $^{+2.2\%}_{-2.9\%}{}^{+7.1\%}_{-7.9\%}$\\[1mm]
 500 & $2.16\times10^{-3}$ & $^{+2.3\%}_{-3.2\%}{}^{+5.5\%}_{-6.8\%}$
     & $2.77\times10^{-3}$ & $^{+2.3\%}_{-3.2\%}{}^{+6.5\%}_{-8.8\%}$
     & $1.10\times10^{-3}$ & $^{+2.5\%}_{-3.3\%}{}^{+7.6\%}_{-8.5\%}$\\[1mm]
 550 & $1.38\times10^{-3}$ & $^{+2.6\%}_{-3.5\%}{}^{+5.8\%}_{-7.3\%}$
     & $1.79\times10^{-3}$ & $^{+2.6\%}_{-3.6\%}{}^{+7.1\%}_{-9.5\%}$
     & $6.93\times10^{-4}$ & $^{+2.7\%}_{-3.6\%}{}^{+8.2\%}_{-9.0\%}$\\[1mm]
 \hline
 \end{tabular}
\caption{The same as TABLE~\ref{tab:CS7}, but for the LHC with
 $\sqrt{s}=14$~TeV}\label{tab:CS14} 
\end{table}
%


\end{document}